\documentclass[numbib,nofootinbib,superscriptaddress,a4paper,twocolumn]{revtex4-1}
\usepackage{geometry}
\usepackage[english]{babel}
\usepackage[latin1]{inputenc}
\usepackage[linktocpage=true]{hyperref}
\usepackage{amsmath, amssymb, amsfonts, mathrsfs}
\usepackage{graphicx}
\usepackage{braket}
\usepackage{makecell}
\usepackage{enumitem}
\usepackage{xspace}
\usepackage{booktabs}

\usepackage{array,mathtools,amssymb}
\newcolumntype{C}{>{$}c<{$}}
\AtBeginDocument{
\heavyrulewidth=.08em
\lightrulewidth=.05em
\cmidrulewidth=.03em
\belowrulesep=.65ex
\belowbottomsep=0pt
\aboverulesep=.4ex
\abovetopsep=0pt
\cmidrulesep=\doublerulesep
\cmidrulekern=.5em
\defaultaddspace=.5em
}

\DeclareRobustCommand{\SkipTocEntry}[1]{} 
\newcommand{\melvin}{{\small M}{\scriptsize ELVIN}\xspace}
\newcommand{\cnot}{{\small C}{\scriptsize NOT}\xspace}

\DeclareRobustCommand{\SkipTocEntry}[5]{}

\begin{document}

\title{Computer-inspired Quantum Experiments}

\author{Mario Krenn}
\email{mario.krenn@univie.ac.at}
\affiliation{Vienna Center for Quantum Science \& Technology (VCQ), Faculty of Physics, University of Vienna, Austria.}
\affiliation{Institute for Quantum Optics and Quantum Information (IQOQI) Vienna, Austrian Academy of Sciences, Austria.}
\affiliation{Department of Chemistry \& Computer Science, University of Toronto, Canada.}
\affiliation{Vector Institute for Artificial Intelligence, Toronto, Canada.}
\author{Manuel Erhard}
\email{manuel.erhard@univie.ac.at}
\affiliation{Vienna Center for Quantum Science \& Technology (VCQ), Faculty of Physics, University of Vienna, Austria.}
\affiliation{Institute for Quantum Optics and Quantum Information (IQOQI) Vienna, Austrian Academy of Sciences, Austria.}
\author{Anton Zeilinger}
\email{anton.zeilinger@univie.ac.at}
\affiliation{Vienna Center for Quantum Science \& Technology (VCQ), Faculty of Physics, University of Vienna, Austria.}
\affiliation{Institute for Quantum Optics and Quantum Information (IQOQI) Vienna, Austrian Academy of Sciences, Austria.}

\begin{abstract}
The design of new devices and experiments in science and engineering has historically relied on the intuitions of human experts. This credo, however, has changed. In many disciplines, computer-inspired design processes, also known as inverse-design, have augmented the capability of scientists. Here we visit different fields of physics in which computer-inspired designs are applied. We will meet vastly diverse computational approaches based on topological optimization, evolutionary strategies, deep learning, reinforcement learning or automated reasoning. Then we draw our attention specifically on quantum physics. In the quest for designing new quantum experiments, we face two challenges: First, quantum phenomena are unintuitive. Second, the number of possible configurations of quantum experiments explodes combinatorially. To overcome these challenges, physicists began to use algorithms for computer-designed quantum experiments. We focus on the most mature and \textit{practical} approaches that scientists used to find new complex quantum experiments, which experimentalists subsequently have realized in the laboratories. The underlying idea is a highly-efficient topological search, which allows for scientific interpretability. In that way, some of the computer-designs have led to the discovery of new scientific concepts and ideas -- demonstrating how computer algorithm can genuinely contribute to science by providing unexpected inspirations. We discuss several extensions and alternatives based on optimization and machine learning techniques, with the potential of accelerating the discovery of practical computer-inspired experiments or concepts in the future. Finally, we discuss what we can learn from the different approaches in the fields of physics, and raise several fascinating possibilities for future research.
\end{abstract}
\maketitle

\flushbottom
\maketitle

\tableofcontents
\section{Introduction}
Computers have long been an indispensable tool for scientists, which enabled far more complex calculations or simulations than possible by humans. While a computer was little more than a calculator traditionally, using hard-coded algorithms predefined by the scientists, this view has changed significantly in the recent past. More and more, computers have been employed in the more \textit{creat}ive design processes, in various fields of physics and thereby augment the conventional intuition-based design strategies. We will first overview which approaches and objectives those fields take. Then we focus our attention on quantum experiments, in particular in the optical regime. Finally, we will examine the similarities and differences between the approaches and objectives and discuss what one can learn from computer-inspired quantum experiments.

\setlength{\tabcolsep}{1pt}
\begin{table*}[t]
\centering
  \begin{tabular}{  c  c c   c c  }
    \toprule
    \textbf{\large Discipline} & \textbf{\large Objectives} & \makecell{\textbf{\large Degrees of}\\ \textbf{\large Freedom}} & \textbf{\large Approaches} & \textbf{\large Status}\\   
    \midrule

    \textbf{Plasma Physics} &  \makecell{magnetic confinement for \\thermonuclear reactor\cite{fasoli2016computational}}  & \makecell{geometry of\\magnetic coils} & \makecell{high-quality\\plasma simulations} & \makecell{Wendelstein 7X\cite{wolf2017major}}\\
    \\
    
    \makecell{\textbf{Accelerator}\\\textbf{Physics}} &  \makecell{stable high-power \\ high-density beams\cite{hofler2013innovative}}  & \makecell{continuous\\ magnet settings \&\\ cavity geometries} & \makecell{evolutionary \&\\  gradient-based} & \makecell{efficient designs\cite{appel2019optimization,pierrick2019klystron}, \\ PoC exp.\cite{li2018genetic}}\\
    \\
    \makecell{\textbf{Mechanical}\\\textbf{Engineering}} &  \makecell{2d \& 3d\\ material designs\cite{bendsoe1988generating,xie1993simple}}  & \makecell{quasi-continuous\\pixel/voxel} & \makecell{topological\\optimization\cite{bentley1999evolutionary, bendsoe2009topology, van2013level, sigmund2011usefulness}} & \makecell{widely used in\\industry \& academia}\\
    \\
    \textbf{Nanophotonics} &  \makecell{nonlinear optics,\\metamaterial,\\topological photonics}  & \makecell{quasi-continuous\\pixel/voxel} & \makecell{topological\\optimization\cite{molesky2018inverse, yao2019intelligent}} & \makecell{efficient designs, \\PoC exp.}\\
\\
    \textbf{Quantum Circuits} &  \makecell{circuits for binary,\\gate-based, universal\\quantum computers}  & \makecell{circuit topology \&\\ parametrized gates} & \makecell{deterministic\cite{bocharov2015efficient, nam2018automated},\\  variational\cite{peruzzo2014variational, farhi2014quantum, mcclean2016theory},\\ machine learning\cite{sim2019expressibility, rattew2019domain},\\automated reasoning\cite{wille2013compact, meuli2018sat, wille2019mapping}} & \makecell{efficient designs, \\PoC exp.\cite{peruzzo2014variational}}\\
\\
    \makecell{\textbf{Superconducting}\\\textbf{Hardware}} &  \makecell{fundamental \\multi-qubit gates\cite{menke2019automated}}  & \makecell{circuit topology \&\\ parametrized gates} & \makecell{topology search,\\ gradient-based} & \makecell{efficient designs}\\
\\
    \makecell{\textbf{Chemistry}} & \makecell{functional materials,\\drug candidates\cite{sanchez2018inverse,gromski2019explore}}  & \makecell{discrete\\molecular graphs} & \makecell{virtual screening\cite{robbins2011simple,gomez2016design, lyu2019ultra},\\ evolutionary\cite{o2011computational, chen2013hybrid, jensen2019graph, nigam2019augmenting} \\ gradient-based\cite{gomez2018automatic}} & \makecell{widely used in\\ industry \& academia}\\
\\
    \makecell{\textbf{Quantum}\\ \textbf{Experiments}} &  \makecell{complex entanglement,\\ transformations in\\quantum optics}  & \makecell{discrete\\experimental topology \& \\ continuous components} & \makecell{highly-efficient\\topological search\cite{krenn2016automated},\\evolutionary\cite{knott2016search, o2019hybrid, nichols2019designing},\\reinforcement\cite{melnikov2018active,wallnofer2019machine},\\gradient-based\cite{arrazola2018machine,adler2019Quantum}} & \makecell{PoC exp.\cite{malik2016multi, schlederer2016cyclic, wang2017generation, babazadeh2017high, erhard2018experimental, kysela2019experimental,xiao2019observation,zhanexperimental},\\conceptual\\ insights\cite{krenn2017entanglement,gao2019computer}}\\

    \bottomrule
  \end{tabular}
  \caption{Computer-inspired designs in selected fields of physics. \textit{PoC exp.} are proof-of-principle experiments.}
\label{tab:ComparisonAll}  
\end{table*}

\subsection{Computer-inspired designs in physics}
One of the most impressive and influential examples of computer-inspired designs can be found in plasma physics, in particular in the engineering of nuclear fusion reactors. A leading concept for creating controlled thermonuclear fusion is based on the magnetic confinement of a plasma in a toric shape. The concept of Tokamaks (symmetrically shaped magnetic coils surrounding the torus for shaping the plasma) has been experimentally studied since the late 1950s; it is the basis of the ITER (International Thermonuclear Experimental Reactor) megaproject. An alternative concept called Stellarator (complexly shaped external magnetic coils for creating ring-shaped, twisted magnet fields) was conceptualised around the same time. However, it has suffered from the inability to precisely shape the magnetic field configurations, which has led to poor experimental performances and subsequent fading of research interest. Only since the late 1980s, when computers became powerful enough to calculate and simulate high-precision magnetic configurations and design the geometry of coil shapes, Stellarators became actively researched again \cite{fasoli2016computational}. The most extensive experimental implementation is the billion-dollar reactor Wendelstein 7X in Greifswald, Germany, with 50 computer-designed, superconducting magnetic coils with a diameter of roughly 3,5 meters. The objective was to design magnetic fields that optimise confinement, stability, transport and equilibrium properties as well as experimental constraints \cite{helander2014theory,pedersen2016confirmation}. The actual coils are subsequently calculated by inverting the resulting field geometries. Wendelstein 7X produced the first stable hydrogen plasma in 2015 \cite{wolf2017major}.

In accelerator physics, computer-inspired and computer-optimised designs have a long tradition \cite{hofler2013innovative}, from which we mention a few interesting recent examples. Genetic algorithms augmented with machine learning techniques have been used to optimise the magnetic confinement configurations of National Synchrotron Light Source II (NSLS-II) Storage Ring (Brookhaven National Laboratory, New York, USA). The goal was to maximise the dynamic aperture, a complex multi-objective function that involves, among others the beam lifetime and energy acceptance and several high-quality solutions have been experimentally tested \cite{li2018genetic}. Quality improvements of the beam injection at the heavy-ion synchrotron SIS18 at GSI/FAIR (Darmstadt, Germany) has also been achieved using genetic algorithms. The algorithm finds a better set of parameters than previous simulation studies \cite{appel2019optimization}. High-power radio-frequency sources, which are essential for particle accelerators, have been optimised using evolutionary algorithms. There, the evolutionary algorithm's objective is to maximise the efficiency of a Klystron, by exploring the continuous  geometric design parameters of a multi-cell cavity \cite{pierrick2019klystron}.

In the field of mechanical engineering, new shapes have been designed by computer algorithms since at least the late 1980s \cite{bendsoe1988generating,xie1993simple}. The field, denoted as topological optimisation \cite{bentley1999evolutionary, bendsoe2009topology, van2013level, sigmund2011usefulness}, approaches structural optimisation problems by discretisation of the two- or three-dimensional space into pixels or voxels (volume pixels), respectively. The individual elements are then manipulated independently by algorithms. In a remarkable recent study, new aerodynamic structures (wings of planes) have been designed using computer-aided approaches \cite{aage2017giga}, to minimise the mass of the structure while keeping the mechanical objectives such as stiffness large. As a result, the new wings could lead to a significant reduction in fuel consumption for aeroplanes.

The idea of topological optimisation has subsequently found application in the field of nano-photonics \cite{molesky2018inverse, yao2019intelligent}, and has since been augmenting the more human-centric intuition-based optimisation schemes. Its applications range from nonlinear optics to topological photonics and nanoscale optics. Concrete examples involve highly efficient free-space-to-waveguide couplers \cite{shen2014integrated}, compact micrometer-scale wavelength demultiplexers for several different colors \cite{su2017inverse}, highly efficient, diamond-based coupling devices \cite{dory2019inverse} or on-chip particle accelerators \cite{sapra2020chip}.

Automated design and verification of logical circuits have long traditions \cite{sheeran2000checking, saeedi2013synthesis}. Its quantum version, the automated synthesis of quantum circuits, is strongly influenced by computer algorithms as well. In principle, the question of designing a quantum circuit for a given quantum algorithm can be solved with the Solovay-Kitaev algorithm for a universal qubit gate sets \cite{dawson2005solovay}. Unfortunately, the algorithm leads to unfeasible large quantum circuits. Thus great efforts are invested in simplifying and optimising quantum circuits with deterministic or heuristic methods \cite{maslov2008quantum, bocharov2015efficient, nam2018automated}. Furthermore, optimising theoretical quantum circuits for hardware with architecture-specific constraints is denoted as compilation \cite{martinez2016compiling,maslov2017basic}. A very different approach is variational optimisation of quantum circuits, using hybrid quantum-classical algorithms. Two prominent examples are VQE (variational quantum eigensolver) \cite{peruzzo2014variational, mcclean2016theory} and QAOA (quantum approximate optimization algorithm) \cite{farhi2014quantum}. There, a parametrised quantum circuit is optimised concerning an objective function. The objective could be the fidelities quantum state or process. Recently identified challenges are exponentially vanishing gradients \cite{mcclean2018barren}, which shows that finding sufficiently good parameter settings requires more unconventional learning algorithms. Furthermore, finding good initial quantum circuit topologies (which is more coarse-grained than in the nano-photonics case) is an active research question \cite{sim2019expressibility, rattew2019domain}. A fundamentally different approach is the application of \textit{automated reasoning}, a field in artificial intelligence that has seen tremendous progress in recent years \cite{brakensiek2019resolution}. The idea is to translate the problem of quantum circuit synthesis to Boolean satisfiability systems. Solutions to the resulting propositional formula can subsequently be found using highly-efficient SAT solvers \cite{wille2013compact, meuli2018sat, wille2019mapping}. 

A recent extension in the family of computer-inspired designs are superconducting quantum hardware \cite{menke2019automated}. In this study, the first efficient, noise-insensitive coupler for four-qubit interactions was discovered, an essential element for quantum simulations. The task is both discrete (defining the topology of the circuit from a space of roughly $10^3$ possibilities) and continuous (setting the parameters of the individual elements in a circuit), which the authors solve by first identifying initial random guesses that are subsequently optimised through gradient descent and swarm optimisation.

In chemistry, computer-inspired design of molecules is widely used for discovering new drugs, functional materials or chemical reactions \cite{sanchez2018inverse,gromski2019explore}. A particular challenge is the enormous search space which is estimated to be in the order of $10^{60}$ even for small bio-molecules \cite{virshup2013stochastic}, and without the possibility of continuous parameters that could be optimised through gradient-based methods. As a result, unbiased and systematic search methods (high-throughput virtual screening) \cite{robbins2011simple,gomez2016design, lyu2019ultra} or genetic algorithms and particle swarms \cite{o2011computational, chen2013hybrid, jensen2019graph, nigam2019augmenting} are a common tool. Very recently, it was shown how these discrete search problems could be solved by transforming them into continuous optimisation problems \cite{gomez2018automatic} -- which opened up the application for deep learning methods in molecular design. 

Computer-inspired designs based on evolutionary strategies for complex, multi-objective problems are used in many other fields of science and engineering \cite{coello2007evolutionary}. A fascinating, massive collection with more than 10.000 relevant references can be found in ref.\cite{coello2010list}.

\section{Computer-inspired quantum experiments}
Our focus here is on quantum experiments, in particular, quantum optics experiments. They have several distinct properties. First, quantum optical experiments consist of discrete optical elements, which makes continuous, gradient-based optimization infeasible in most situations. Second, the number of possible experimental configurations is enormous, and even for relatively small setups can go beyond $10^{20}$ possible settings. Third, even for quite small quantum states or quantum transformations, it can be very challenging to find suitable experimental configurations using just intuition. Nonetheless, most of the time, experimental setups have been, and still are designed by human quantum physicists.

We will cover general high-dimensional multi-particle quantum optical systems \cite{pan2012multiphoton,flamini2018photonic}. While the experimental technology has shown impressive improvements over the last years in generating \cite{graffitti2017pure,lenzini2017active,wang201818,luo2019quantum}, manipulating\cite{bornman2019ghost,wang2019boson,hu2019experimental} and measuring \cite{bavaresco2018measurements,ahn2019adaptive} complex entangled quantum systems, often the lack of feasible experimental design proposals hinders further progress. 

\begin{figure}[t]
\centering
\includegraphics[width=0.98\linewidth]{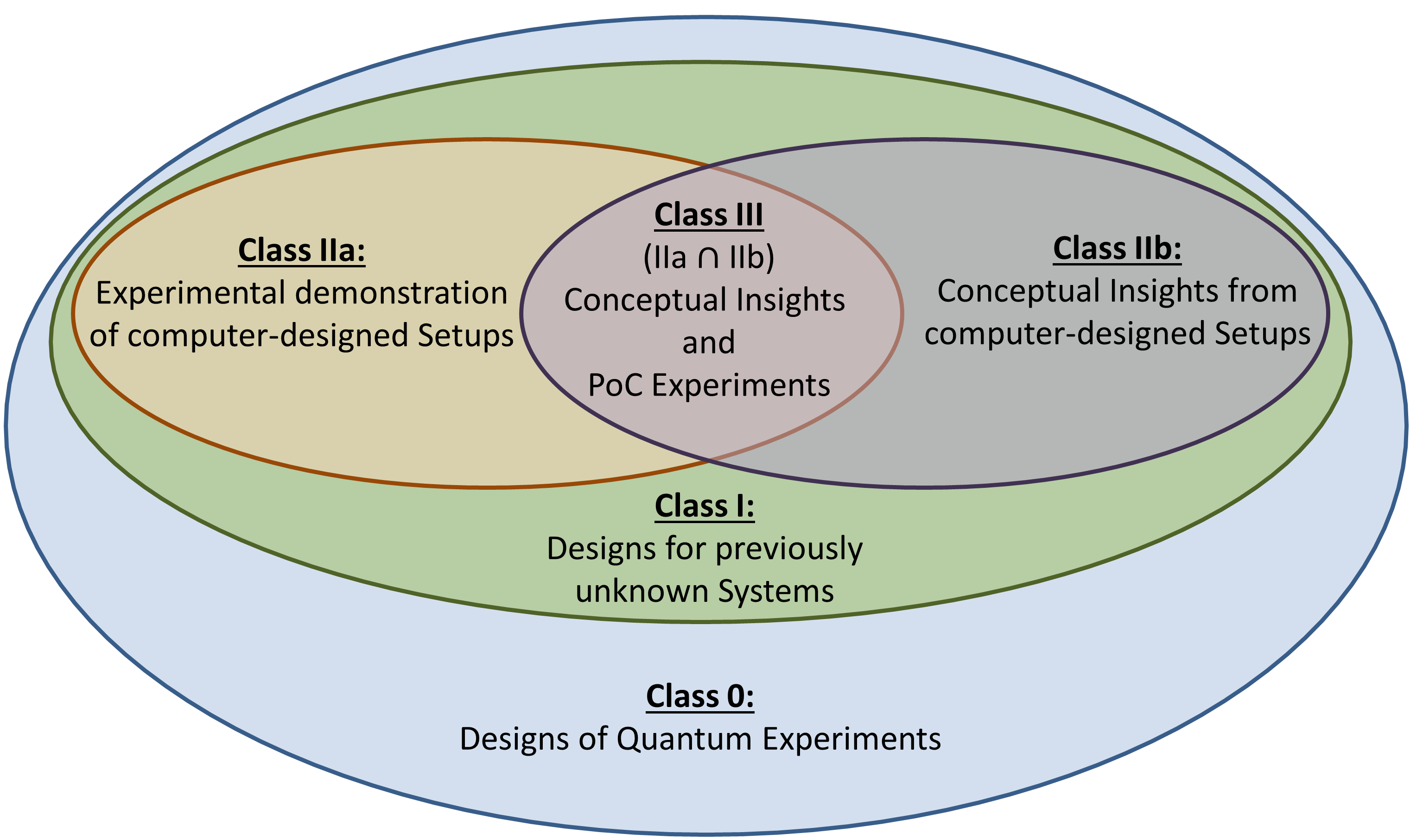}
\caption{A categorisation of algorithms for designing quantum experiments. Algorithms in a higher class have shown to be applicable in more practical ways of contributing to scientific research. The details about different classes are explained in the main text.}
\label{fig:hierarchy}
\end{figure}

We will focus on \textit{practical} alternative to the human-centred design of quantum experiments. For that, we define \textbf{classes of practicality} of algorithms for computer-designed quantum experiments, which indicate the level of maturity and demonstrated applicability in the scientific domain:
\begin{enumerate}[align=left]
\item[\textbf{Class 0:}] The algorithm has rediscovered solutions to previously solved questions.

\item[\textbf{Class I:}] The algorithm has uncovered solutions to previously unsolved questions.

\item[\textbf{Class IIa:}] The algorithm has uncovered solutions to previously unsolved questions, which have been experimentally demonstrated.

\item[\textbf{Class IIb:}] The algorithm has inspired the discovery of scientific insights or concepts.

\item[\textbf{Class III:}] Combination of Class IIa and IIb. The algorithm has uncovered solutions to previously unsolved questions, which have been experimentally demonstrated, and has inspired the discovery of scientific insights or concepts.

\end{enumerate}
In Fig. \ref{fig:hierarchy}, we show a graphically the relations of these classes. This classification is sufficient for the moment for our purpose. In the future, however, hopefully much more surprising and far-reaching insights can be obtained from algorithms; therefore, Class IIb should then be more fine-grained.

Next, we are going to detail different approaches. First, we start with an algorithm for efficient topological search which lies in class III. Afterwards, we describe a method that has led to new experimental designs that have been implemented in laboratories, which lies in class IIa. Finally, we overview several other promising techniques that aim to design new quantum experiments in class I. Those techniques are mainly based on optimization and machine-learning techniques and have the potential to be of practical influence in the future.

Afterwards, we will connect computer-inspired quantum experiments with inverse-design techniques in other fields of physics, and thereby indicate several interdisciplinary ideas that might be interesting future research questions.

\subsection{Class III: Highly-efficient topological search for practical designs and discovery of scientific concepts}\label{melvinSection}
We will describe a computational strategy that can be used for automated design of new quantum experiments and fulfil all three properties. The underlying idea has been presented in 2016, denoted as \melvin\cite{krenn2016automated}, and has since been significantly extended and improved. It has led to solutions of several previously unsolved questions  \cite{krenn2016automated, gao2019arbitrary}, many computer-designed experiments have been experimentally implemented \cite{malik2016multi, schlederer2016cyclic, wang2017generation, babazadeh2017high, erhard2018experimental, kysela2019experimental} and it has been used to discover new scientific ideas and concepts \cite{krenn2017entanglement, krenn2017quantum,gao2019computer}.

The program can be considered as a highly efficient and optimized search routine for the inherently discrete topology of quantum optical setups\footnote{The calculation of quantum states is based on symbolic transformations. Example codes both for Wolfram Mathematica as well as for Python (using SymPy) can be found at \href{https://github.com/XuemeiGu/MelvinPython/}{github.com/XuemeiGu/MelvinPython/}}. While continuous gradient-based optimization techniques are efficient in identifying (local) optima, discrete topological search has the great advantage that their results are often interpretable for human scientists. That is because quantum states in the resulting experimental setup mostly consist of only a small number of terms, the associated probabilities are low-order rational numbers. As a consequence, when a solution is identified, it is often possible for the human to grasp the underlying principle.

While a topological search is conventionally inefficient, we developed two core ideas which significantly accelerates the identification of solutions:
\begin{enumerate}  
\item Generalized objective functions to increase the probability of finding potential solutions
\item Identification of necessary criteria which allow to abort calculations before evaluating time-expensive properties
\end{enumerate}  
Because of the algorithm's practicality, we will demonstrate underlying principles on two examples and show how it can inspire not only new experimental configurations but also new conceptual ideas.

\subsubsection{Concrete Example: 3-dimensional Greenberger-Horne-Zeilinger state}\label{lab3dGHZ}

\begin{figure}[t]
\centering
\includegraphics[width=0.90\linewidth]{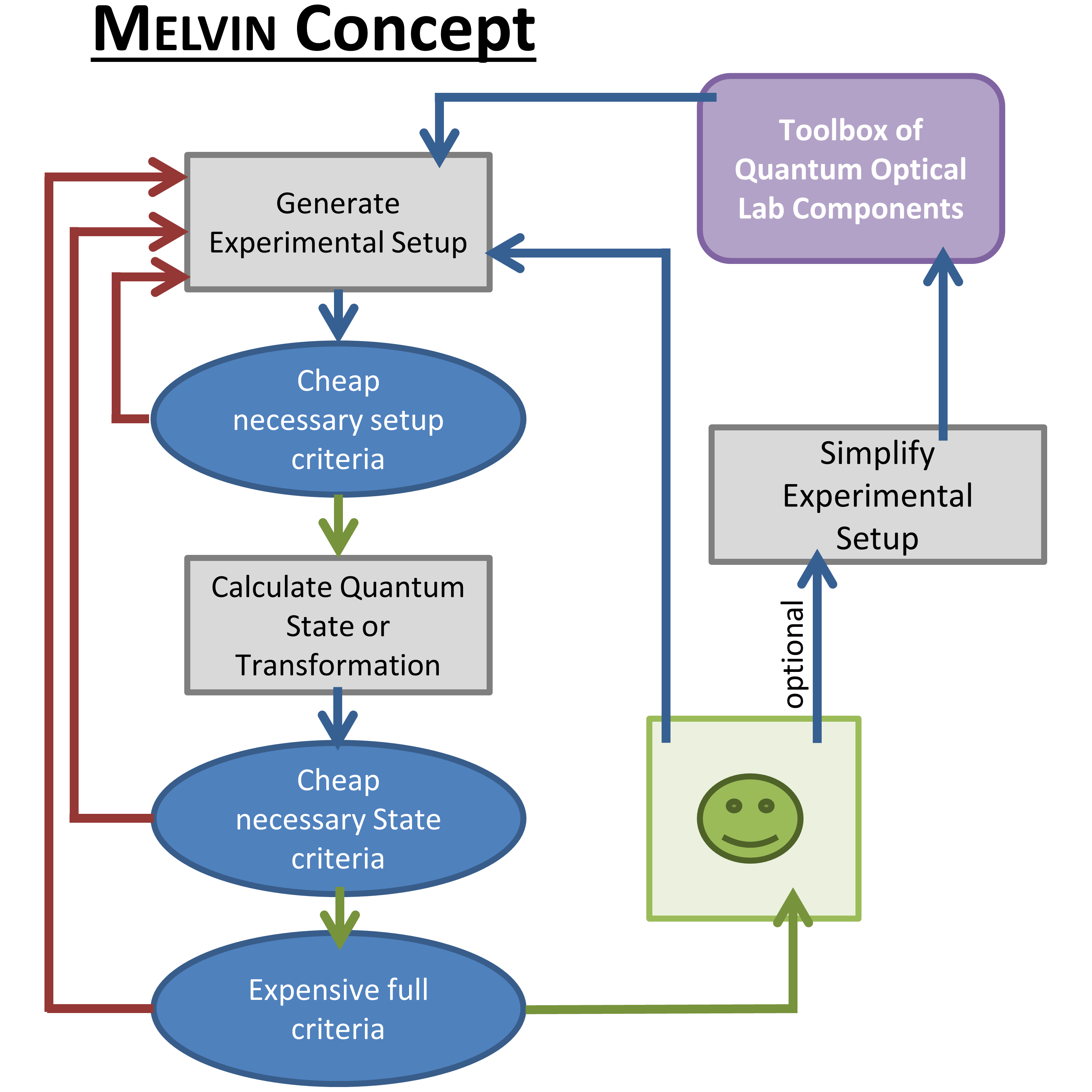}
\caption{Concept of the Class-III algorithm for computer-inspired experiments, \melvin, based on ref.\cite{krenn2016automated}. Experiments are assembled from a toolbox that contains all optical elements available in the laboratory. Time-inexpensive criteria applied before calculating the full quantum state. Afterwards, again, time-inexpensive criteria are applied before calculating the full, usually expensive objective. If the experiment fulfils all criteria, it is reported to the user -- otherwise, a new setup is generated. Optionally, the setup is simplified and appended to the toolbox, such that it can be used in subsequent trials to generate more complex solutions quicker. If the low-cost criteria and the expensive objective function are defined broadly, the search routine is highly efficient. We show how to do this in the main text.}
\label{melvinconcept}
\end{figure}

The principle idea of the algorithm for designing quantum experiments is shown in Fig. \ref{melvinconcept}, and we will discuss practical considerations for finding new interesting experimental ideas and novel concepts in quantum physics, based on concrete examples. 

\begin{figure}[t]
\centering
\includegraphics[width=0.85\linewidth]{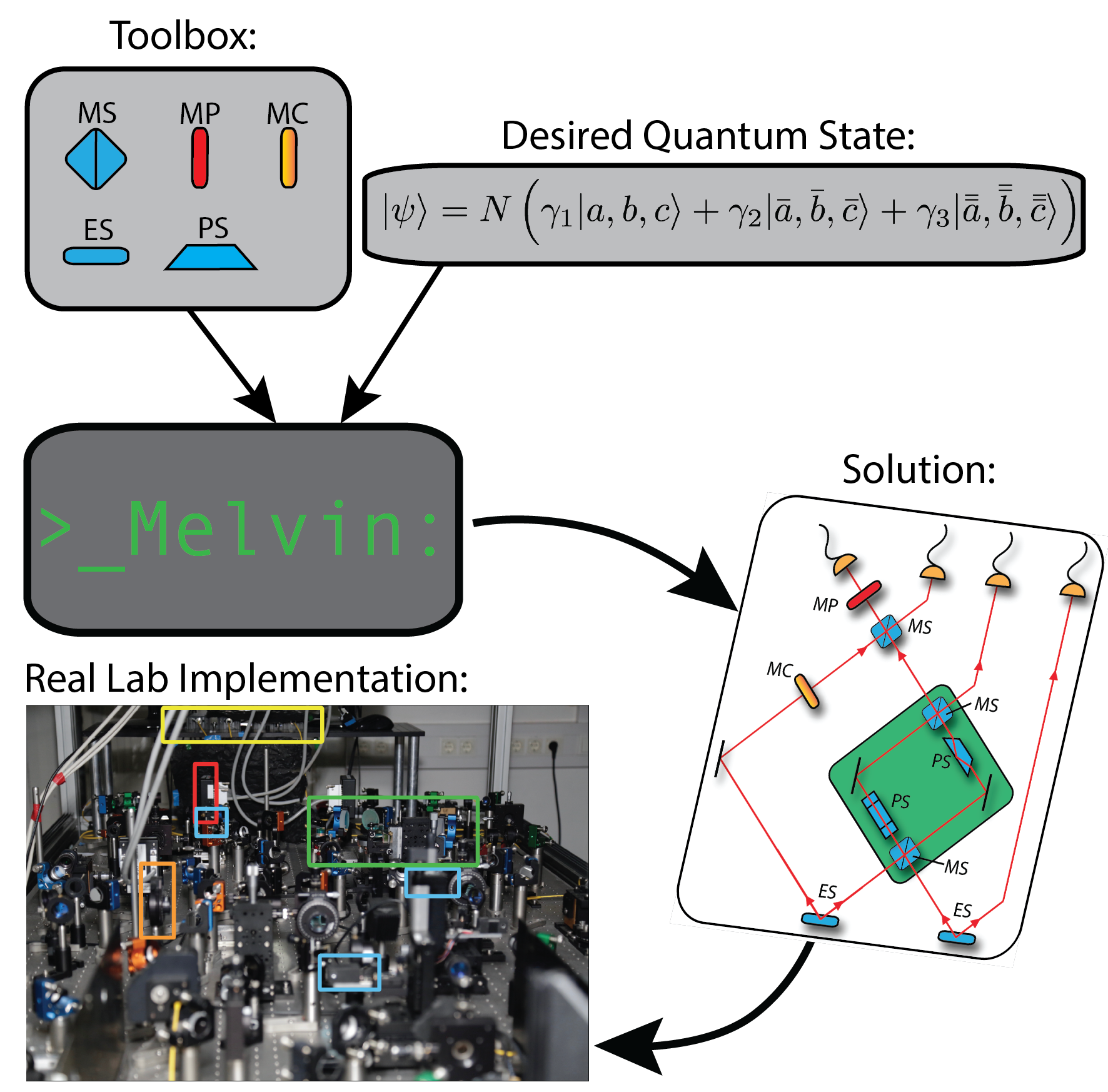}
\caption{Intuition about complexity of computer-inspired quantum experiments. The algorithm discovered the experimental setup for a 3-dimensional GHZ state only using the Toolbox and the generalized quantum state as inputs. The toolbox contains, mode-splitter (MS), mode-projectors (MP), mode-changers (MC), phase-shifters (PS) and entanglement-sources (ES). Human scientists can then implement the concept in the laboratory. It allows for the successful experimental creation and investigation of the target state\cite{erhard2018experimental}. }
\label{figureGHZ}
\end{figure}

We will explain the application on the 3-dimensional 3-partite Greenberger-Horne-Zeilinger (GHZ) state, which was the first example presented in ref. \cite{krenn2016automated}, and has been experimentally demonstrated (with some intermediate steps \cite{malik2016multi}) more than three years later \cite{erhard2018experimental}. An intuition of the complexity of a computer-designed experiment can be seen in Fig.\ref{figureGHZ}. The initial discovery of the 3d GHZ setup in ref.\cite{krenn2016automated} required roughly 5 CPU-core hours with a Intel Core i5-2540 2.5GHz, and could be significantly accelerated with methods presented now.

The three-dimensional GHZ state can be written as
\begin{align}
	\ket{\psi}=\frac{1}{\sqrt{3}}\left(\ket{0,0,0}+\ket{1,1,1}+\ket{2,2,2}\right),
  \label{3dghz}
\end{align}
where $\ket{0}$, $\ket{1}$ and $\ket{2}$ are two orthogonal states, independent of their degree of freedom.

\textbf{Assembling a Quantum Experiment:}\label{sec:toolbox} The algorithms starts by placing two spontaneous parametric down-conversion (SPDC) crystals on virtual optical table. The rest of the setups is assembled from elements in the toolbox. The toolbox contains optical devices which are accessible in a quantum optics lab. Examples are beam splitters, phase shifters, wave plates, holograms, Dove prisms. 
In addition to elementary transformations, adding composite devices to the toolbox can significantly speed up the search process. Those are specially useful transformations in quantum optics, such as effective single-photon filters via quantum teleportation \cite{wang2015quantum}, local high-dimensional gate transformations \cite{babazadeh2017high}, special stable polarisation transformations \cite{anwer2019experimental}. In our example, the introduction of an interferometer-based parity sorter for spatial modes of photons (introduced by Leach et al. \cite{leach2002measuring}) has been shown to decrease the search time by a factor of roughly 25 \cite{krenn2016automated}.

After a certain number of element placed on the virtual optical table, the resulting state is calculated. For the 3-dimensional GHZ state, one of the four particle is used as a trigger to herald the generation of a three-photon state in the other three detectors. The resulting quantum state is then compared to the search target.

\textbf{Generalized objective functions: }\label{melvinobjective} Here we show, on a concrete example, how to define very general objectives, which can immensely accelerate the search process. The main challenge is that the search space can be enormous. For example, if six photonic modes are used, and the toolbox consists of two 2-input-2-output elements (each has 6$\cdot$5=30 possible locations), and 10 single-input elements (such as holograms with different mode numbers, or prisms with different discrete angles; each of could be in a certain path), one has 120 different choices of elements. For a standard optical experiment with 15 different elements, one results in roughly $120^{15}\approx10^{31}$ different configurations. The idea is to generalize the objective function as much as possible, such that the number of correctly identified quantum experiments is as large as possible and thereby the possibility of identifying a useful solution is maximal.

The state in equation eq.(\ref{3dghz}) is very specific, and a search for only this state is very narrow. To increase the chances of satisfying the objective, one needs to \textit{formulate the target state in the most general way}. In particular, here every local unitary transformation results in a 3-dimensional GHZ state
\begin{align}
	\ket{\psi}=\frac{1}{\sqrt{3}}\left(\ket{a,b,c}+e^{i\phi_1}\ket{\bar{a},\bar{b},\bar{c}}+e^{i\phi_2}\ket{\bar{\bar{a}},\bar{\bar{b}},\bar{\bar{c}}}\right),
  \label{3dghz2}
\end{align}
with $x\perp\bar{x}\perp\bar{\bar{x}}$ and $x \in \{a,b,c\}$, and arbitrary phases $\phi_1$ and $\phi_2$. Even under very conservative assumptions\footnote{We assume an encoding space for the modes from $\ket{-2}$ to $\ket{2}$, and eight different possible phase values phase steps in $e^{i \pi/4}$, and later we will different phase steps. Experimentally, smaller steps are feasible.}, these leads to more than 2 million times more potential targets than the natural objective in eq.(\ref{3dghz}). In addition, while changing the magnitude of the coefficients of the three terms changes the structure of entanglement, equal coefficients could easily be recovered just by experimentally feasible, local mode-dependent filters, which leads to the following objective state
\begin{align}
	\ket{\psi}=N\left(\gamma_1\ket{a,b,c}+\gamma_2\ket{\bar{a},\bar{b},\bar{c}}+\gamma_3\ket{\bar{\bar{a}},\bar{\bar{b}},\bar{\bar{c}}}\right),
  \label{3dghz3}
\end{align}
with $\gamma_i \in \mathbb{C}$, and $N$ being a normalisation constant. Under similar conservative assumptions, this leads to more than $5\cdot10^7$ times as many targets as objective eq.(\ref{3dghz}). Importantly, if we find any of these targets, we immediately know how to generate experimentally a 3-dimensionally entangled GHZ state with Fidelity $F=1$.

One can even further generalise the objective when considering that local mode-dependent filters are experimentally simple operations. Therefore, it is conceivable to reformulate the objective into
\begin{align}
	&\ket{\psi}=N\Large(\gamma_1\ket{a,b,c}+\gamma_2\ket{\bar{a},\bar{b},\bar{c}}+\gamma_3\ket{\bar{\bar{a}},\bar{\bar{b}},\bar{\bar{c}}}+\nonumber\\
	&\sum_{\ell_1,\ell_2} \gamma_{1,\ell_1,\ell_2}\ket{\tilde{a},\ell_1,\ell_2}+\gamma_{2,\ell_1,\ell_2}\ket{\ell_1,\tilde{b},\ell_2}+\gamma_{3,\ell_1,\ell_2}\ket{\ell_1,\ell_2,\tilde{c}}\Large),
  \label{3dghz4}
\end{align}
with $\tilde{x}$ being orthogonal to $x$, $\bar{x}$ and $\bar{\bar{x}}$. This leads to a significantly larger number of targets states recognized as successful experimental suggestion of the objective in eq.(\ref{3dghz4}), and for any of these solutions, it is immediately clear how to generate the state with fidelity $F=1$. This enormously larger target space translates into a significant speed up in the search procedure, and indicate the great importance of the definition of a general objective function. \footnote{The speed up is not linearly related to the increased target space, because physical constraints experimental devices will not cover the full space.}

Alternatively instead of using a specific target state as an objective, one could try to find a specific, general property of interest -- which is a more general concept. In the current example, one can ask \textit{Is the entanglement shown in eq.(\ref{3dghz}) the only acceptable structure, or could other types of high-dimensional states also lead to interesting solutions?}. As an example, one could ask whether the states 
\begin{align}
	\ket{\psi_1}=\frac{1}{\sqrt{3}}\left(\ket{0,0,0}+\ket{1,1,1}+\ket{2,2,1}\right),\nonumber\\
	\ket{\psi_2}=\frac{1}{2}\left(\ket{0,0,0}+\ket{1,0,1}+\ket{2,1,0}+\ket{3,1,1}\right),
  \label{3dghz5}
\end{align}
also be considered as interesting solutions? If the answer is yes, one can select states according to generalized entangled properties. In this example, a concept denoted as \textit{Schmidt-Rank Vector} (SRV) introduced by Huber and de Vicente \cite{huber2013structure,huber2013entropy} perfectly fits the task. The SRV generalises the concept of high-dimensional entanglement to multiple particles, and denotes a vector of the dimensionalities of entanglement of every biparition. Loosely speaking, it shows the dimensionality of entanglement between one particle and the rest of the quantum state. For three particles (which is our case), the SRV of the state in eq. (\ref{3dghz}) is (3,3,3) as every party is three-dimensionally entangled with the remaining state. States in eq.(\ref{3dghz5}) are (3,3,2) and (4,2,2) dimensionally entangled. The SRV gives a new way to classify high-dimensional entangled states. It can be used as the objective (instead of a direct state), which increases the number of possible useful solutions enormously.

\textbf{Speed Up: Identifying low-cost, necessary criteria:} Calculating the final quantum state emerging from the experimental setup as well as evaluating the objective functions can be very time-expensive. For that reason, we want to find efficient methods which tell us whether we should perform the full calculation at all. For our example, one necessary condition for the creation of a genuine multiparty entangled state is that the quantum information from the two initial entangled pairs mix during the evaluation of the setup. This criterion (which is generally applicable to many quantum photonic experimental states) can be applied even before the quantum state has been calculated, by investigating the structure of the generated experimental setup. If the experiment does not fulfil the criterion, it can be discarded without calculation.

After calculating the quantum state, and before evaluating the full objective function of state in eq.(\ref{3dghz3}) or eq.(\ref{3dghz4}), or the Schmidt-Rank Vector (which involves the time-expensive calculation of density matrix ranks), one can analyse whether the calculated state has at least three different modes in each in each of the three photons. This criteria significantly reduces the number of states which undergo the full criteria evaluation.

\subsubsection{Generalized objectives: High-Dimensional Quantum Gates}
We give one more example of a generalized objective function, which has been successfully applied in the identification of high-dimensional multi-photon quantum gates \cite{gao2019computer}. We aim to write the objective in the most general, non-trivial way, which allows for the largest number of potential useful solutions. The simplest case is a controlled operation with two control modes and three target modes:
\begin{align}
	\textnormal{\cnot}\ket{0,0}=\ket{0,0}, \textnormal{\cnot}\ket{1,0}=\ket{1,1},\nonumber\\ 
  \textnormal{\cnot}\ket{0,1}=\ket{0,1}, \textnormal{\cnot}\ket{1,1}=\ket{1,2},\nonumber\\
  \textnormal{\cnot}\ket{0,2}=\ket{0,2}, \textnormal{\cnot}\ket{1,2}=\ket{1,0}.
  \label{2d-cnot}
\end{align}
At this point, we need to extract the essence of the transformation in an objective function. All mode numbers can be general, but there are much more degrees of freedom, which can significantly improve the chances of success.

After many attempts using different objective functions, the following has lead to the first successful solution,
\begin{align}
	\textnormal{\cnot}\ket{c_1,t_{1,2,3}}=\ket{x_{1,2,3},\bar{t}_{1,2,3}},\nonumber\\
	\textnormal{\cnot}\ket{c_2,t_{1,2,3}}=\ket{y_{1,2,3},\bar{\bar{t}}_{1,2,3}},
  \label{2d-cnot2}
\end{align}
where $c_{1/2}$ stand for the two different control modes, $t_{1/2/3}$ are the three target modes. $x$ and $y$ stand for the two output control modes. Conventionally they are unchanged, but allowing the freedom of change increases enormously the success probability. Furthermore, $\bar{t}$ and $\bar{\bar{t}}$ are target output modes, which satisfy the following criteria $\bar{t_1}\perp\bar{t_2}\perp\bar{t_3}$, $\bar{\bar{t_1}}\perp\bar{\bar{t_2}}\perp\bar{\bar{t_3}}$ and $\bar{t_1}\perp\bar{\bar{t_1}}$, $\bar{t_2}\perp\bar{\bar{t_2}}$ and $\bar{t_3}\perp\bar{\bar{t_3}}$. For every proposed setup, a large list of input control and target modes are used to calculate the output of the setup. If any subset of input modes lead to results that fulfil the criteria in eq.(\ref{2d-cnot2}), a control gate has likely been discovered.

An important key insight is the following: The criteria above are only necessary, not sufficient -- i.e. there can be solutions which are not controlled gate operations. We refrain from defining more strict criteria because from experience, we have seen that \textit{near miss} solutions can be adjusted by human scientists into a full solution. This significantly increases the chances to find inspirations for human scientists.
\begin{figure}[t]
\centering
\includegraphics[width=0.85\linewidth]{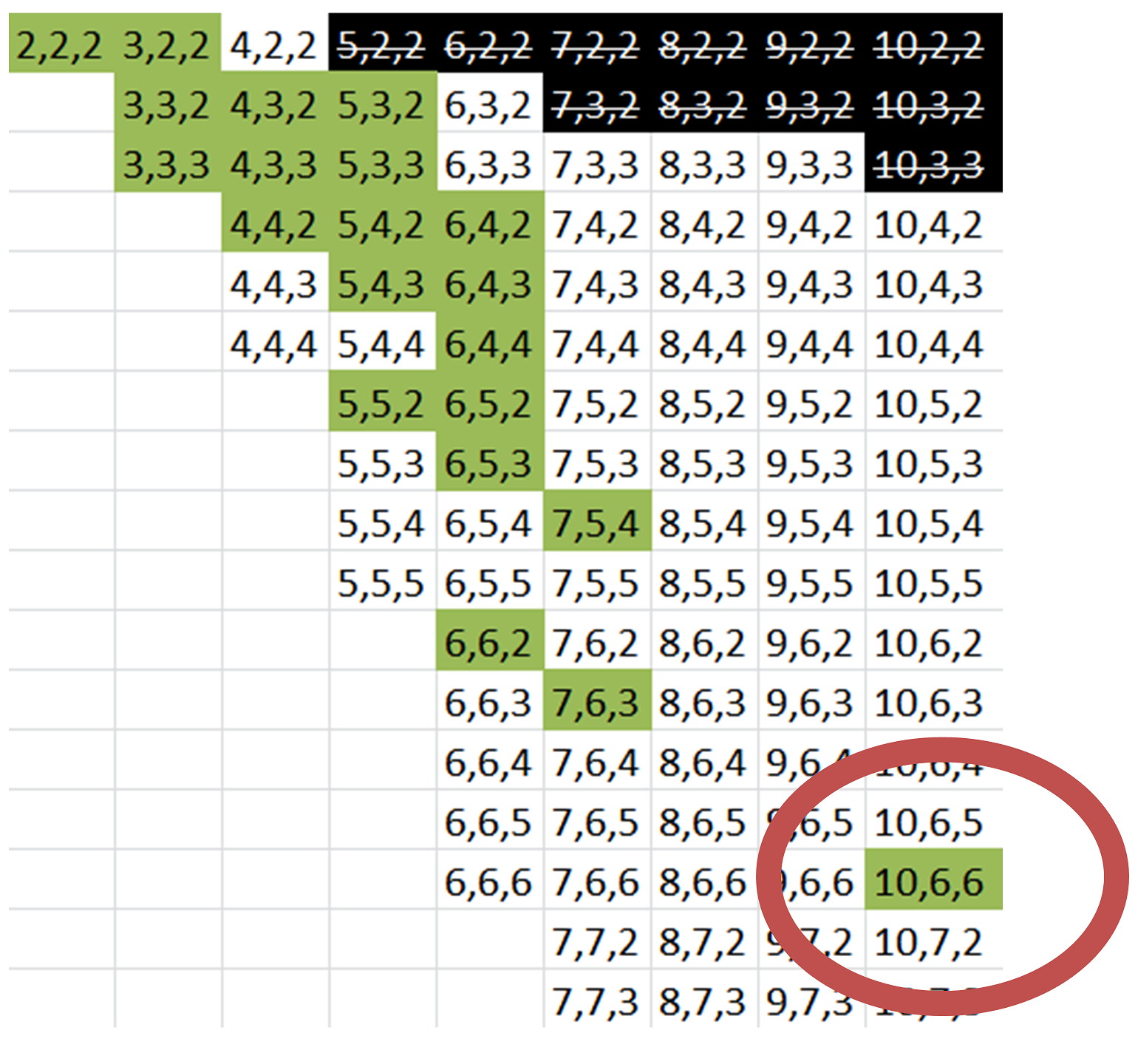}
\caption{A table of Schmidt-Rank vectors for three-photon entangled quantum states, as investigated using \melvin. Green cells stand for cases where solutions have been found. For white cells, no solutions have been found, and black cells show algebraic impossible SRVs \cite{huber2013structure}. The red circle indicates an outlier solution. The investigation of the experimental setup leading to this solution has led to the concept of \textit{Entanglement by Path Identity} \cite{krenn2017entanglement}.}
\label{figureSRVlist}
\end{figure}

Identifying solutions for complex transformations are much more computational expensive; identifying the first high-dimensional control gate transformation has required roughly 150k CPU-core hours in ref.\cite{gao2019computer}.
\subsubsection{Discovery of computer-inspired Concepts}
Finding solutions to a predefined objective can have a significant influence on future research. For example, the solution to the 3-dimensional GHZ state allows now to investigate statements about the local and realistic properties of the universe experimentally \cite{PhysRevA.89.024103,lawrence2014rotational,lawrence2019many}. 
\begin{figure*}[ht]
\centering
\includegraphics[width=0.95\textwidth]{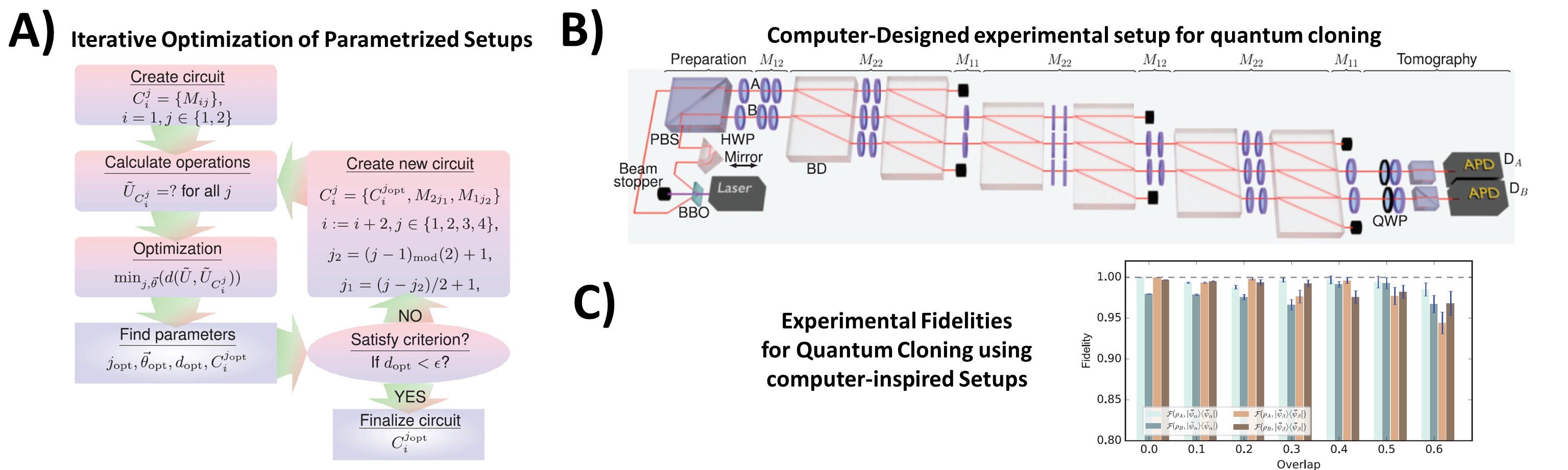}
\caption{Algorithm of Class IIA. \textbf{A)} The concept of the algorithm is to start building a whole setup from continuously parametrized building blocks. The program starts with one building block and optimizes its free parameters (or example, angle settings of half-wave and quarter-wave plates). If the target state cannot be reached within a certain quality, the algorithm adds another basic building blocks with continuous parameters and continues to optimize the angle setting until it reaches the specified target quality. \textbf{B)} A setup for performing deterministic, non-unitary quantum cloning, with three sets of building blocks. BD stands for beam displacer, and are polarization-dependent objects. Blue objects between the BDs are wave-plates, black boxes are loss elements. Discovering this setup required 30 minutes at an Intel Core i5 notebook with 1.6 GHz and 8 GB RAM using Wolfram Mathematica. \textbf{C)} High-quality experimental results for the fidelity of quantum-cloned states. Image from ref.\cite{zhanexperimental}}
\label{experiment_optimizations_peng}
\end{figure*}
Scientifically even more interesting would be the case where we could learn new concepts or ideas from solutions found by computer algorithms. This is possible, as shown in \cite{krenn2017entanglement}, where an entirely new concept for the generation of high-dimensional multi-particle entanglement has been presented. In this case, the key insight was to use the algorithm with an objective that allows a large number of different classes of solutions. In this concrete case, the objective was finding states with different Schmidt-Rank vectors, as described in section \ref{melvinobjective}.

The initial state, in this case, are two SPDC processes that each produce three-dimensionally entangled photon pairs. One natural limit would suggest that the maximally achievable entanglement is $d=3\times 3=9$ dimensional. After running \melvin for roughly 50.000 CPU-hours (with the settings explained in the previous chapter), it produced quantum entangled states with 21 different Schmidt-Rank vector structures, as shown in Fig. \ref{figureSRVlist}. All except for one solution satisfied the natural limit of $d\leq9$, while one solution was a clear outlier, achieving 10-dimensional entanglement.

Investigations of the experimental setup showed that it contained an implicit usage of a technique pioneered by Zou, Wang and Mandel in 1991 \cite{zou1991induced}. The technique was not part of the toolbox, it was not allowed explicitly by the rules of the algorithm, and it was not known to the authors prior to the discovery. Instead, the solution contained a non-local interferometer, which was allowed but not enforced in any way by the human operators. The non-linear interferometer allows for the implicit usage of the Zou-Wang-Mandel technique.

When it was understood that and how Zou-Wang-Mandel's idea can be used in the context of high-dimensional, multipartite entanglement generation, the human scientists were able to generalize it to many different cases, and has since been denoted as \textit{Entanglement by Path Identity} \cite{krenn2017entanglement}. Soon after it was understood, that this type of entanglement generation is closely related to and can efficiently be described by Graph Theory \cite{krenn2017quantum, gu2019quantum2, gu2019quantum3,krenn2019questions}.

The example above shows how computer algorithms can not only find solutions to explicitly defined problems but can also -- accidentally \cite{lehman2018surprising} -- lead to the discovery of new ideas and concepts that maybe would have never been found by human scientists. 

\subsection{Class IIA: Iterative optimization of parametrized, practical setups}
Quantum experiments using the path degree-of-freedom as a carrier of information have seen enormous progress over the last few years \cite{flamini2018photonic,wang2019integrated,fengprogress}. There, efficient algorithms for decomposing high-dimensional single-photon transformations into experimental setups are well known for unitary \cite{reck1994experimental, clements2016optimal} or non-unitary \cite{tischler2018quantum} cases. However, in the presence of more than one photon, or quantum information carried by more than one degree-of-freedom, finding suitable experimental configurations is very challenging, even if one has access to highly controlled setups.

To experimentally investigate processes that depend on non-unitarity, such as dynamics involving PT-symmetry \cite{xiao2019observation} and deterministic quantum cloning exploiting non-unitarity \cite{zhanexperimental}, the group of Xue have relied on computer algorithms for designing their experimental setups, see Fig.\ref{experiment_optimizations_peng}A. The algorithm iteratively increases the experimental setup (in the form of Fig.\ref{experiment_optimizations_peng}B) block by block. Each block consists of eights independent, continuous parameters (settings of half-wave plates and quarter-wave plates), two input and output paths as well as two loss-paths.

The algorithm starts with one block and optimizes the free parameters towards its objective function. If the difference between optimized state and target state is too large, the algorithm extends the setup by another block and continues to optimize the new parameters. This process continues until the algorithm has found a setting with $n$ blocks of 8$n$ parameters which can reach the target function.

Fig.\ref{experiment_optimizations_peng}B represents the topology of a setup with three blocks, which was used in ref.\cite{zhanexperimental} to perform deterministic, non-unitary quantum cloning. The team was able to experimentally implement this complex experimental configuration, resulting in a very high-quality average fidelity of beyond 98\%.

The algorithm has demonstrated its ability to design setups for previously unsolved questions, and its solutions have been practical such that they where experimentally execute in a laboratory, thus it is in the Class IIA . Finally, it will be interesting to understand how new scientific insights and ideas can be extracted from this approach (which would correspond to Class III).

\subsection{Class I: Optimization and machine learning methodologies}

\begin{figure*}[ht]
\centering
\includegraphics[width=0.95\textwidth]{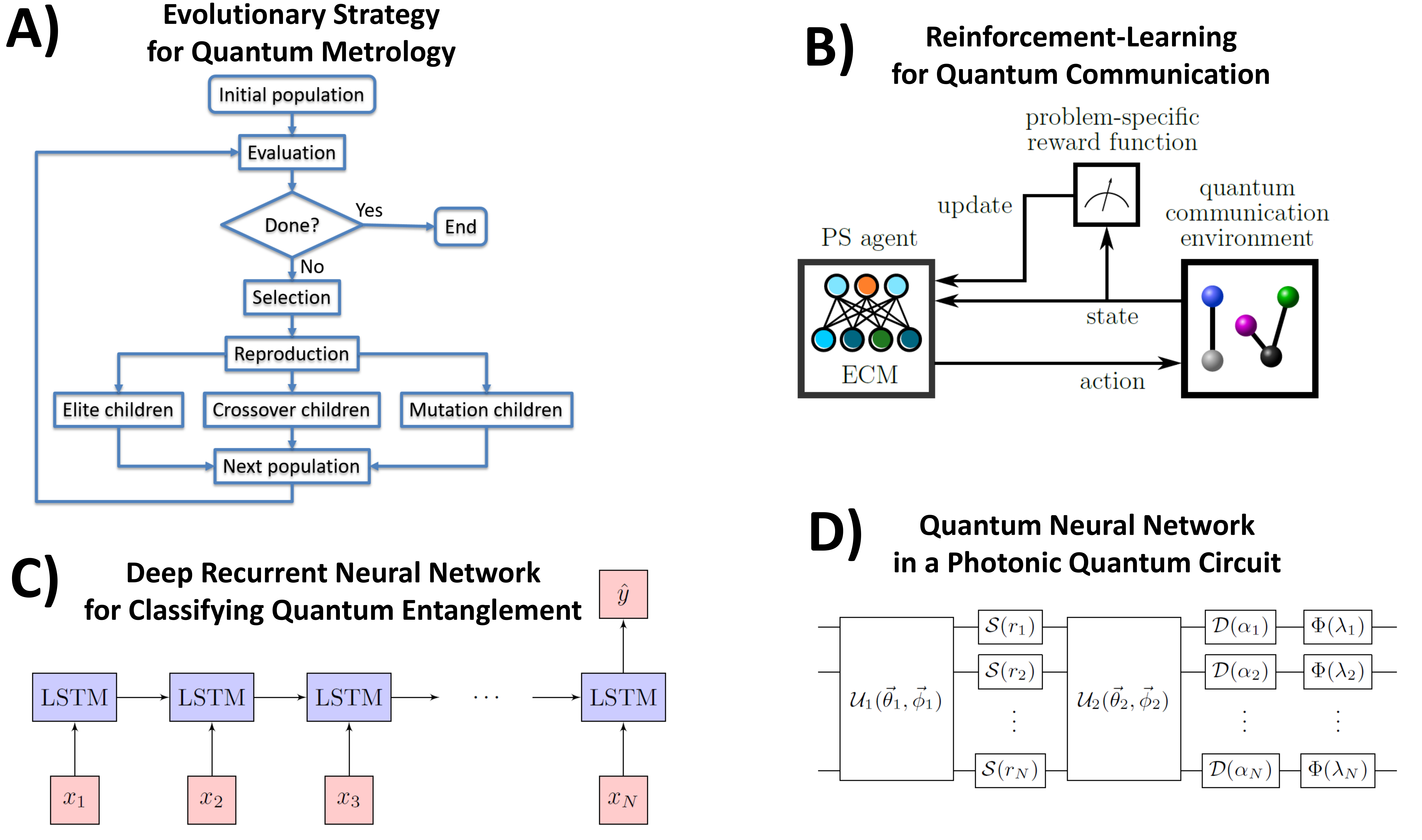}
\caption{Class-I algorithms for computer-inspired quantum experiments with various optimization and machine learning techniques. \textbf{A)} An genetic algorithm for designing novel methods in quantum metrology. An initial, random population undergoes an evolutionary process. The individual setups, which form the population, are selected according to their fitness. The best ones are mutated and form the next generation. Image from ref.\cite{nichols2019designing} \textbf{B)} The concept of a reinforcement learning algorithm for designing quantum communication schemes. An agent performs actions in an environment (changing quantum communication scheme), which changes the state of the environment. The agent receives a reward according to the quality of the action. Image from ref. \cite{wallnofer2019machine}. \textbf{C)} A deep recurrent neural network, based on long-short-term (LSTM) cells, receives optical elements in a sequence ($x_i$), and learns to predict quantum entanglement properties of the setups $\hat{y}$. In that way, the time-consuming objective function in a design process is approximated by a fast neural network, which has the potential to speed up the search for new quantum entanglement experiments significantly. Image from ref. \cite{adler2019Quantum} \textbf{D)} A photonic quantum circuit can be parametrized continuously, thus gradient-based optimization techniques are possible. The circuits topology here follows the \textit{quantum neural network} Ansatz, which consists of unitaries $U_i$, squeezing $S$, displacement $D$ and non-Gaussian gates $\Phi$. Image from ref.\cite{arrazola2018machine}.}
\label{optimizations}
\end{figure*}
The algorithms in the previous chapter have led to experimental designs that were subsequently successfully implemented in laboratories (Class IIA) and have led to new scientific insights (Class III). The approaches in this chapter have not yet demonstrated that level of maturity, which will be an essential question for future research. Nonetheless, they use new optimization and machine learning approaches which could in the future lead to exciting practical developments; thus, we will explain them in detail. We start by mentioning that a simple method to adaptively improve the search algorithm is storing successful experiments as a whole inside the toolbox for later use, as shown in Fig.\ref{melvinconcept}. This has been demonstrated in ref.\cite{krenn2016automated} to improve the performance in special cases by an order of magnitude.

\subsubsection{Evolutionary Strategies}
To design quantum experimental setups for efficient quantum metrology \cite{giovannetti2011advances}, Knott has demonstrated a genetic algorithm approach \cite{knott2016search}. Evolutionary strategies have gained a lot of attention recently in machine learning since it was discovered that they are scalable alternatives to reinforcement algorithms \cite{salimans2017evolution}.

The algorithm \textit{Tachikoma}\footnote{named after the AI-robots in the famous anime Ghost in the Shell} has access to a toolbox of experimentally available elements, in a similar way as described in chapter \ref{sec:toolbox}. However, in contrast to \cite{krenn2016automated}, the toolbox is filled with continuous-variable quantum optical technologies. The experimental elements include squeezing operator, displacement operators, beam splitters with variable transmission rates, phase operator, photon-number sensitive measurements and quadrature measurements. The continuous parameters are further restricted to resemble experimentally feasible operations, such that the final results are practical states.

The algorithm starts with a population of randomly assembled initial experiments, chosen from the toolbox. Those initial solutions then undergo evolutionary optimisation, by mutating the experimental setups and selecting the best candidates as offspring for the next generation. The selection criteria are based on a fitness function that involves the phase-measuring capability for a given number of average photon numbers. Indeed, Tachikoma uncovered several promising candidate setups, that more than doubles the precision compared to the best-known, practical state. 

In the three years since the original publication, the approach has been significantly improved. In one extension, the authors show how the usage of deep neural networks can lead to a speedup in finding useful output state. The idea is to train a neural network to classify the photon number distribution of a given state into one out of six categories (which involves cat states, squeezed cat states, cubic phases and others). The network helps to guide the search in the right direction, and only later, the computational expensive fitness functions are evaluated within the genetic algorithm. Thereby, several useful quantum states have been discovered \cite{o2019hybrid}.

The authors also extend their work of quantum metrology, called \textit{AdaQuantum}\footnote{after Ada Lovelace, the world's first computer programmer, and resident of Nottingham where the algorithm has been designed}. By improving both the numerical simulation and refining the search algorithm, the authors are able to improve the speed by another factor of five over their own best result. Furthermore, they show higher noise and photon loss tolerance, which is essential in real experimental situations \cite{nichols2019designing}, see Fig.\ref{optimizations}A. If these promising proposals are indeed feasible experimentally, and yield the predicted phase measurement precision, they could become prime examples in experimental quantum metrology research.

Those computations were executed for 96 hours on 16 cores of the University of Nottingham's High-Performance Computing facility. Further improvements of the algorithm, as outlined in the text, could enable the optimisation in the significantly larger space of three optical modes. This could not only improve the resulting measurement precision but also allow the algorithm to explore even more unorthodox solutions which physicists could then try to understand. In that spirit, the authors conclude their manuscript by asking \textit{why the experiments we presented here to perform so well in maximising their respective fitness functions. This might be especially revealing in the states that are robust to noise: why do these specific experimental arrangements found by AdaQuantum create states that still perform well with photon loss?}\cite{nichols2019designing}. Finding the underlying conceptual reasons could indeed uncover new insights for human scientists, thus it is an exciting future research direction.

\subsubsection{Reinforcement Learning}
In a reinforcement learning (RL) scenario, an agent observes and takes actions in an environment to maximize some reward function \cite{sutton1998introduction}. The idea became widely known by the impressive results on playing computer games \cite{mnih2015human, vinyals2019grandmaster, jaderberg2019human}, defeat world-champion go and chess players \cite{silver2018general}. Spearheaded by investigations by Schmidhuber and others since the early 1990s, it has been understood how to set up environments such that artificial agents show behaviour that mimics \textit{curiosity}, \textit{creativity} or \textit{intrinsic motivation} \cite{schmidhuber1991curious,schmidhuber1991possibility,schmidhuber2010formal}. Curiosity-driven exploration has very recently been demonstrated in autonomous gameplay learning \cite{pathak2017curiosity, pathak18largescale}. Those results motivate the application of RL techniques to scientific environments, and in particular to the design of new quantum experiments.

In ref.\cite{melnikov2018active}, the authors employed a RL algorithm (denoted as \textit{Projective Simulations} \cite{briegel2012projective,briegel2012creative}) to the environment of quantum experiments and quantum states. The agent acts in the same environment as \melvin\cite{krenn2016automated}, as described in chapter \ref{lab3dGHZ}. Its task is to build an optical experiment, element by element, using a toolbox of optical elements. The reward of the environment depends on the entanglement property of the resulting quantum state. Over time, the agent learns to choose optical elements in such a way that it finds more interesting quantum states. With this approach, the algorithm re-discovers solutions that have been discovered previously by \melvin. Besides, it autonomously finds ways to simplify the experiments, which previously has only been achieved in a hard-coded way. The algorithm's internal representation show that it learned to build specific optical devices that have been used by humans for many years, such as a special type of interferometer\cite{leach2002measuring}. 

After confirming that the RL approach works for re-discovery tasks, a similar algorithm was targeted to identify new quantum communication protocols\cite{wallnofer2019machine}. This is an important step, as the authors correctly say, \textit{We are aware that we make use of existing knowledge in the specific design of the challenges. Rediscovering existing protocols under such guidance is naturally very different from the original achievement (by humans) of conceiving of and proposing them in the first place, an essential part of which includes the identification of relevant concepts and resources} \cite{wallnofer2019machine}. For that reason, the projective simulation RL algorithm \cite{briegel2012projective, melnikov2018active} has subsequently been applied to challenging tasks where solutions are not well studied and understood by human scientists \cite{wallnofer2019machine}. There, the algorithm discovered a quantum repeater that shows better performance in realistic, asymmetrical situations than the best-known previous solution, as shown in Fig.\ref{optimizations}B. With this achievement, the algorithm lies in the Class-I. It would be exciting whether the proposed experimental setup is actually feasible and can be implemented in real-world laboratories, and whether the predicted performance matches experimental results. Furthermore, it would be fascinating to understand which new conceptual insights scientists can extract from the solutions of the agent. Those are important open questions for the future.

\subsubsection{Supervised Classification for Speedup}
Supervised learning methods (for classification or generation tasks) rely on a large amount of data. If such a dataset can be provided, supervised learning methods can be used to model quantum optical experiments and thereby potentially significantly reducing the number of required search steps compared to an entirely unguided search. In ref.\cite{adler2019Quantum}, a deep recurrent neural network in the form of a Long-Short-Term-Memory (LSTM) network \cite{hochreiter1997long} was able to model and predict complex entanglement properties of a state that is the output of a quantum experiment, see Fig.\ref{optimizations}C. The network receives as an input an optical element sequence and is trained to predict the corresponding Schmidt-Rank vector, which has been introduced already in chapter \ref{melvinobjective}. The training data consist of 100.000s of examples, extended from the table in Fig.\ref{figureSRVlist}. The LSTM network shows much better than random prediction qualities, which is a necessary and promising first step towards a deep generative model for quantum experiments based on deep reinforcement learning.

\subsubsection{Gradient-Based Optimization in Quantum Photonic Circuits}
Integrated photonics allow for access to arbitrary unitary operations \cite{reck1994experimental} that are continuously parametrized and can span a large discrete \cite{qiang2018large,wang2018multidimensional,lu2019three,fengprogress,wang2019integrated} or continuous \cite{weedbrook2012gaussian,lenzini2018integrated,zhang2019integrated} Hilbert spaces.

Such systems are ideally suited for gradient-based optimization of quantum states and quantum transformations. In a work by Arrazola et al. \cite{arrazola2018machine}, the authors aim to find optimal quantum circuit settings to produce useful quantum systems for continuous-variable (CV) quantum architectures. They use a variational quantum circuit called \textit{CV quantum neural networks}, whose gates and connectivities (i.e. its topology) are fixed, but where all the gates contain free parameters \cite{killoran2018continuous}, see Fig.\ref{optimizations}D. They find implementations for important states such as NOON states, which are essential in quantum metrology \cite{giovannetti2011advances} and Gottesman-Kitaev-Preskill states, which can be used for error correction in CV quantum computing \cite{menicucci2014fault}. In addition, they task their algorithm (which is based on the Strawberry Fields software platform \cite{killoran2019strawberry}) with finding implementations for cubic phase gates, which is a crucial basis element for CV quantum computation \cite{sabapathy2019production}, as well as the famous Quantum Fourier Transformation. All of their results have theoretical fidelity larger than 99\%. It would be exciting to see whether those promising solutions that are in theory of significantly higher quality than previously known ones are feasible to be implemented experimentally and whether extracting scientific understanding from those high-quality solutions is possible.

\section{Where to go from here?}
One of the most unique and essential characteristics of computer-designed quantum experiments in Class IIB or Class III (as \melvin in section \ref{melvinSection}) is the possibility of extracting inspirations for conceptual and scientific understanding. We argue that this was made possible because of the pure topological search. While most other approaches involve, at least partly, the optimisation of continuous parameters, computer-inspired quantum experiments are designed mainly through course-grained topological optimisation. As a consequence, solutions contain clearly identifiable patterns, and the resulting quantum states are sparse. Those properties facilitate interpretability \cite{krenn2017entanglement, gao2019arbitrary, gao2019computer}. It would be interesting to apply a similar approach also in other fields, especially in the computer-inspired designs of quantum circuits and superconducting hardware \cite{menke2019automated}. While a purely topological search is likely significantly slower in identifying optimal solutions, it might lead to more interpretable solutions. An interesting related challenge is to understand how to improve interpretability in solutions of continuous optimisation strategies.

The objectives researched for quantum optical experiments could potentially directly be investigated through nanophotonics \cite{molesky2018inverse}. For example, one could envision a nanophotonic structure, which directly emits high-dimensional multiphotonic entangled states. Alternatively, a high-efficient and stable generalisation of computer-designed holographic transformations \cite{morizur2010programmable, fontaine2019laguerre, brandt2019high} or scattering \cite{rotter2017light, fickler2017custom, leedumrongwatthanakun2019programmable} to multi-photonic states could potentially significantly reduce experimental complexities in bulk optics.

The field of computer-inspired molecules in chemistry has very similar questions as those raised here. There, discrete objects (atoms) form an enormous search space of $10^{40}$-$10^{60}$ possibilities, even for relatively small molecules. It was only very recently when it has been shown how this discrete optimisation problem can be formulated in a continuous manner \cite{gomez2018automatic}. Continuous optimisation allows for the exploitation of gradient descent, and thus the application of modern deep learning methods, which has become a vivid field both in academia and in industry \cite{sanchez2018inverse, gromski2019explore}. Quantum experiments can be interpreted as graphs in a similar way as molecules \cite{krenn2019selfies}. Thus most artificial intelligence and high-throughput technologies from chemistry could directly be applied to the design of novel quantum setups.

Topological search and verification of electric circuits have long been employing automated reasoning technologies, which is a pure logic-based artificial intelligence technique, which has seen remarkable progress over the last decade \cite{heule2016solving,heule2017science,brakensiek2019resolution}. Similar techniques have also been explored for gate-based qubit quantum circuits \cite{wille2013compact, meuli2018sat, wille2019mapping}. Reformulating search for topologies of quantum optical experiments as propositional formula would be a fascinating field of research.

Furthermore, it would be useful to find ways how the algorithm's results and internal representation of the problem can be easier interpreted physically and intuitively. This could be very helpful for humans trying to understand and learn new concepts and design rules from the discovered solutions. One example could be interpretable neural networks \cite{higgins2017beta,chen2018isolating} in the physical context  \cite{lusch2018deep,iten2020discovering,Nautrup2020operationally} that are applied on deep generative models for complex scientific structures such as functional molecules or quantum experiments \cite{krenn2019selfies}.

Another critical question is how unexpected solutions can be identified more systematically? They have the potential to stimulate new creative insights which humans have not thought of yet \cite{bentley2002introduction, lehman2018surprising}, and are particularly desirable in the scientific context. 

Interestingly, algorithms, as outlined in this manuscript, can not only be applied to quantum experiments themselves but have already been used to find solutions for questions in theoretical quantum information \cite{pavivcic2019automated}. Many problems in the foundations of entanglement theory are of discrete nature \cite{ goyeneche2015absolutely,bengtsson2017geometry,horodecki2020five}, which makes them a great target to apply methods described here. Solutions to questions in theoretical quantum information (rather than experimental designs) might be directly interpretable and lead to conceptual insights.

Even partial answers to these question could lead to new, exciting seeds of computer-inspired ideas. In the best case, such algorithms will become tools to augment human scientist's creativity. We think that this is possible.

\addtocontents{toc}{\SkipTocEntry}
\section*{Acknowledgements}
This work was supported by the Austrian Academy of Sciences ({\"O}AW), University of Vienna via the project QUESS and the Austrian Science Fund (FWF) with SFB F40 (FOQUS). ME acknowledges support from FWF project W 1210-N25 (CoQuS). MK acknowledges support from the FWF via the Erwin Schr\"odinger fellowship No. J4309. 

\newpage

\let\oldaddcontentsline\addcontentsline
\renewcommand{\addcontentsline}[3]{}
\bibliographystyle{unsrt}
\bibliography{sample}

\begin{thebibliography}{}

\bibitem{fasoli2016computational}
A. Fasoli, S. Brunner, W. Cooper, J. Graves, P. Ricci, O. Sauter and L.
  Villard, Computational challenges in magnetic-confinement fusion physics.
  \textit{Nature Physics} \textbf{12}, 411 (2016).

\bibitem{wolf2017major}
R. Wolf, A. Ali, A. Alonso, J. Baldzuhn, C. Beidler, M. Beurskens, C.
  Biedermann, H.S. Bosch, S. Bozhenkov, R. Brakel and  others, Major results
  from the first plasma campaign of the Wendelstein 7-X stellarator.
  \textit{Nuclear Fusion} \textbf{57}, 102020 (2017).

\bibitem{hofler2013innovative}
A. Hofler, B. Terzi{\'c}, M. Kramer, A. Zvezdin, V. Morozov, Y. Roblin, F. Lin
  and C. Jarvis, Innovative applications of genetic algorithms to problems in
  accelerator physics. \textit{Physical Review Special Topics-Accelerators and
  Beams} \textbf{16}, 010101 (2013).

\bibitem{appel2019optimization}
S. Appel, W. Geithner, S. Reimann, M. Sapinski, R. Singh and D. Vilsmeier,
  Optimization of Heavy-Ion Synchrotrons Using Nature-Inspired Algorithms and
  Machine Learning. \textit{13th Int. Computational Accelerator Physics
  Conf.(ICAP'18), Key West, FL, USA, 20-24 October 2018} 15--21 (2019).

\bibitem{pierrick2019klystron}
H. Pierrick, P. Juliette, M. Claude and P. Franck, Klystron efficiency
  optimization based on a genetic algorithm. \textit{2019 International Vacuum
  Electronics Conference (IVEC)} 1--2 (2019).

\bibitem{li2018genetic}
Y. Li, W. Cheng, L.H. Yu and R. Rainer, Genetic algorithm enhanced by machine
  learning in dynamic aperture optimization. \textit{Physical Review
  Accelerators and Beams} \textbf{21}, 054601 (2018).

\bibitem{bendsoe1988generating}
M.P. Bends{\o}e and N. Kikuchi, Generating optimal topologies in structural
  design using a homogenization method. \textit{Computer methods in applied
  mechanics and engineering} \textbf{71}, 197--224 (1988).

\bibitem{xie1993simple}
Y.M. Xie and G.P. Steven, A simple evolutionary procedure for structural
  optimization. \textit{Computers \& structures} \textbf{49}, 885--896 (1993).

\bibitem{bentley1999evolutionary}
P. Bentley, Evolutionary design by computers. (Morgan Kaufmann, 1999).

\bibitem{bendsoe2009topology}
M.P. Bends{\o}e, Topology optimization. (Springer, 2009).

\bibitem{van2013level}
N.P. Dijk, K. Maute, M. Langelaar and F. Van~Keulen, Level-set methods for
  structural topology optimization: a review. \textit{Structural and
  Multidisciplinary Optimization} \textbf{48}, 437--472 (2013).

\bibitem{sigmund2011usefulness}
O. Sigmund, On the usefulness of non-gradient approaches in topology
  optimization. \textit{Structural and Multidisciplinary Optimization}
  \textbf{43}, 589--596 (2011).

\bibitem{molesky2018inverse}
S. Molesky, Z. Lin, A.Y. Piggott, W. Jin, J. Vuckovi{\'c} and A.W. Rodriguez,
  Inverse design in nanophotonics. \textit{Nature Photonics} \textbf{12},
  659--670 (2018).

\bibitem{yao2019intelligent}
K. Yao, R. Unni and Y. Zheng, Intelligent nanophotonics: merging photonics and
  artificial intelligence at the nanoscale. \textit{Nanophotonics} \textbf{8},
  339--366 (2019).

\bibitem{bocharov2015efficient}
A. Bocharov, M. Roetteler and K.M. Svore, Efficient synthesis of probabilistic
  quantum circuits with fallback. \textit{Physical Review A} \textbf{91},
  052317 (2015).

\bibitem{nam2018automated}
Y. Nam, N.J. Ross, Y. Su, A.M. Childs and D. Maslov, Automated optimization of
  large quantum circuits with continuous parameters. \textit{npj Quantum
  Information} \textbf{4}, 23 (2018).

\bibitem{peruzzo2014variational}
A. Peruzzo, J. McClean, P. Shadbolt, M.H. Yung, X.Q. Zhou, P.J. Love, A.
  Aspuru-Guzik and J.L. O'brien, A variational eigenvalue solver on a photonic
  quantum processor. \textit{Nature communications} \textbf{5}, 4213 (2014).

\bibitem{farhi2014quantum}
E. Farhi, J. Goldstone and S. Gutmann, A quantum approximate optimization
  algorithm. \textit{arXiv:1411.4028} (2014).

\bibitem{mcclean2016theory}
J.R. McClean, J. Romero, R. Babbush and A. Aspuru-Guzik, The theory of
  variational hybrid quantum-classical algorithms. \textit{New Journal of
  Physics} \textbf{18}, 023023 (2016).

\bibitem{sim2019expressibility}
S. Sim, P.D. Johnson and A. Aspuru-Guzik, Expressibility and entangling
  capability of parameterized quantum circuits for hybrid quantum-classical
  algorithms. \textit{arXiv:1905.10876} (2019).

\bibitem{rattew2019domain}
A.G. Rattew, S. Hu, M. Pistoia, R. Chen and S. Wood, A Domain-agnostic,
  Noise-resistant Evolutionary Variational Quantum Eigensolver for
  Hardware-efficient Optimization in the Hilbert Space.
  \textit{arXiv:1910.09694} (2019).

\bibitem{wille2013compact}
R. Wille, N. Przigoda and R. Drechsler, A compact and efficient SAT encoding
  for quantum circuits. \textit{2013 Africon} 1--6 (2013).

\bibitem{meuli2018sat}
G. Meuli, M. Soeken and G. De~Micheli, Sat-based $\{$CNOT, T$\}$ quantum
  circuit synthesis. \textit{International Conference on Reversible
  Computation} 175--188 (2018).

\bibitem{wille2019mapping}
R. Wille, L. Burgholzer and A. Zulehner, Mapping quantum circuits to IBM QX
  architectures using the minimal number of SWAP and H operations.
  \textit{Proceedings of the 56th Annual Design Automation Conference 2019} 142
  (2019).

\bibitem{menke2019automated}
T. Menke, F. H{\"a}se, S. Gustavsson, A.J. Kerman, W.D. Oliver and A.
  Aspuru-Guzik, Automated discovery of superconducting circuits and its
  application to 4-local coupler design. \textit{arXiv preprint
  arXiv:1912.03322} (2019).

\bibitem{sanchez2018inverse}
B. Sanchez-Lengeling and A. Aspuru-Guzik, Inverse molecular design using
  machine learning: Generative models for matter engineering. \textit{Science}
  \textbf{361}, 360--365 (2018).

\bibitem{gromski2019explore}
P.S. Gromski, A.B. Henson, J.M. Granda and L. Cronin, How to explore chemical
  space using algorithms and automation. \textit{Nature Reviews Chemistry} 1
  (2019).

\bibitem{robbins2011simple}
D.W. Robbins and J.F. Hartwig, A simple, multidimensional approach to
  high-throughput discovery of catalytic reactions. \textit{Science}
  \textbf{333}, 1423--1427 (2011).

\bibitem{gomez2016design}
R. G{\'o}mez-Bombarelli, J. Aguilera-Iparraguirre, T.D. Hirzel, D. Duvenaud, D.
  Maclaurin, M.A. Blood-Forsythe, H.S. Chae, M. Einzinger, D.G. Ha, T. Wu and
  others, Design of efficient molecular organic light-emitting diodes by a
  high-throughput virtual screening and experimental approach. \textit{Nature
  materials} \textbf{15}, 1120 (2016).

\bibitem{lyu2019ultra}
J. Lyu, S. Wang, T.E. Balius, I. Singh, A. Levit, Y.S. Moroz, M.J. O'Meara, T.
  Che, E. Algaa, K. Tolmachova and  others, Ultra-large library docking for
  discovering new chemotypes. \textit{Nature} \textbf{566}, 224 (2019).

\bibitem{o2011computational}
N.M. O'Boyle, C.M. Campbell and G.R. Hutchison, Computational design and
  selection of optimal organic photovoltaic materials. \textit{The Journal of
  Physical Chemistry C} \textbf{115}, 16200--16210 (2011).

\bibitem{chen2013hybrid}
X. Chen, W. Du, R. Qi, F. Qian and H. Tianfield, Hybrid gradient particle swarm
  optimization for dynamic optimization problems of chemical processes.
  \textit{Asia-Pacific Journal of Chemical Engineering} \textbf{8}, 708--720
  (2013).

\bibitem{jensen2019graph}
J.H. Jensen, A graph-based genetic algorithm and generative model/Monte Carlo
  tree search for the exploration of chemical space. \textit{Chemical science}
  \textbf{10}, 3567--3572 (2019).

\bibitem{nigam2019augmenting}
A. Nigam, P. Friederich, M. Krenn and A. Aspuru-Guzik, Augmenting Genetic
  Algorithms with Deep Neural Networks for Exploring the Chemical Space.
  \textit{arXiv:1909.11655} (2019).

\bibitem{gomez2018automatic}
R. G{\'o}mez-Bombarelli, J.N. Wei, D. Duvenaud, J.M. Hern{\'a}ndez-Lobato, B.
  S{\'a}nchez-Lengeling, D. Sheberla, J. Aguilera-Iparraguirre, T.D. Hirzel,
  R.P. Adams and A. Aspuru-Guzik, Automatic chemical design using a data-driven
  continuous representation of molecules. \textit{ACS central science}
  \textbf{4}, 268--276 (2018).

\bibitem{krenn2016automated}
M. Krenn, M. Malik, R. Fickler, R. Lapkiewicz and A. Zeilinger, Automated
  search for new quantum experiments. \textit{Physical Review Letters}
  \textbf{116}, 090405 (2016).

\bibitem{knott2016search}
P. Knott, A search algorithm for quantum state engineering and metrology.
  \textit{New Journal of Physics} \textbf{18}, 073033 (2016).

\bibitem{o2019hybrid}
L. O'Driscoll, R. Nichols and P. Knott, A hybrid machine learning algorithm for
  designing quantum experiments. \textit{Quantum Machine Intelligence}
  \textbf{1}, 5--15 (2019).

\bibitem{nichols2019designing}
R. Nichols, L. Mineh, J. Rubio, J.C. Matthews and P.A. Knott, Designing quantum
  experiments with a genetic algorithm. \textit{Quantum Science and Technology}
  \textbf{4}, 045012 (2019).

\bibitem{melnikov2018active}
A.A. Melnikov, H.P. Nautrup, M. Krenn, V. Dunjko, M. Tiersch, A. Zeilinger and
  H.J. Briegel, Active learning machine learns to create new quantum
  experiments. \textit{Proceedings of the National Academy of Sciences}
  \textbf{115}, 1221--1226 (2018).

\bibitem{wallnofer2019machine}
J. Walln{\"o}fer, A.A. Melnikov, W. D{\"u}r and H.J. Briegel, Machine learning
  for long-distance quantum communication. \textit{arXiv:1904.10797} (2019).

\bibitem{arrazola2018machine}
J.M. Arrazola, T.R. Bromley, J. Izaac, C.R. Myers, K. Br{\'a}dler and N.
  Killoran, Machine learning method for state preparation and gate synthesis on
  photonic quantum computers. \textit{Quantum Science and Technology} (2018).

\bibitem{adler2019Quantum}
T. Adler, M. Erhard, M. Krenn, J. Brandstetter, J. Kofler and S. Hochreiter,
  Quantum Optical Experiments Modeled by Long Short-Term Memory.
  \textit{arXiv:1910.13804} (2019).

\bibitem{malik2016multi}
M. Malik, M. Erhard, M. Huber, M. Krenn, R. Fickler and A. Zeilinger,
  Multi-photon entanglement in high dimensions. \textit{Nature Photonics}
  \textbf{10}, 248 (2016).

\bibitem{schlederer2016cyclic}
F. Schlederer, M. Krenn, R. Fickler, M. Malik and A. Zeilinger, Cyclic
  transformation of orbital angular momentum modes. \textit{New Journal of
  Physics} \textbf{18}, 043019 (2016).

\bibitem{wang2017generation}
F. Wang, M. Erhard, A. Babazadeh, M. Malik, M. Krenn and A. Zeilinger,
  Generation of the complete four-dimensional Bell basis. \textit{Optica}
  \textbf{4}, 1462--1467 (2017).

\bibitem{babazadeh2017high}
A. Babazadeh, M. Erhard, F. Wang, M. Malik, R. Nouroozi, M. Krenn and A.
  Zeilinger, High-dimensional single-photon quantum gates: Concepts and
  experiments. \textit{Physical Review Letters} \textbf{119}, 180510 (2017).

\bibitem{erhard2018experimental}
M. Erhard, M. Malik, M. Krenn and A. Zeilinger, Experimental
  Greenberger--Horne--Zeilinger entanglement beyond qubits. \textit{Nature
  Photonics} \textbf{12}, 759 (2018).

\bibitem{kysela2019experimental}
J. Kysela, M. Erhard, A. Hochrainer, M. Krenn and A. Zeilinger, Experimental
  High-Dimensional Entanglement by Path Identity. \textit{arXiv:1904.07851}
  (2019).

\bibitem{xiao2019observation}
L. Xiao, K. Wang, X. Zhan, Z. Bian, K. Kawabata, M. Ueda, W. Yi and P. Xue,
  Observation of critical phenomena in parity-time-symmetric quantum dynamics.
  \textit{Physical Review Letters} \textbf{123}, 230401 (2019).

\bibitem{zhanexperimental}
X. Zhan, K. Wang, L. Xiao, Z. Bian, Y. Zhang, B.C. Sanders, C. Zhang and P.
  Xue, Experimental quantum cloning in a pseudo-unitary system.
  \textit{Physical Review A} \textbf{101}, R010302 (2020).

\bibitem{krenn2017entanglement}
M. Krenn, A. Hochrainer, M. Lahiri and A. Zeilinger, Entanglement by path
  identity. \textit{Physical Review Letters} \textbf{118}, 080401 (2017).

\bibitem{gao2019computer}
X. Gao, M. Erhard, A. Zeilinger and M. Krenn, Computer-inspired concept for
  high-dimensional multipartite quantum gates. \textit{arXiv:1910.05677}
  (2019).

\bibitem{helander2014theory}
P. Helander, Theory of plasma confinement in non-axisymmetric magnetic fields.
  \textit{Reports on Progress in Physics} \textbf{77}, 087001 (2014).

\bibitem{pedersen2016confirmation}
T.S. Pedersen, M. Otte, S. Lazerson, P. Helander, S. Bozhenkov, C. Biedermann,
  T. Klinger, R.C. Wolf and H.S. Bosch, Confirmation of the topology of the
  Wendelstein 7-X magnetic field to better than 1: 100,000. \textit{Nature
  communications} \textbf{7}, 13493 (2016).

\bibitem{aage2017giga}
N. Aage, E. Andreassen, B.S. Lazarov and O. Sigmund, Giga-voxel computational
  morphogenesis for structural design. \textit{Nature} \textbf{550}, 84 (2017).

\bibitem{shen2014integrated}
B. Shen, P. Wang, R. Polson and R. Menon, Integrated metamaterials for
  efficient and compact free-space-to-waveguide coupling. \textit{Optics
  express} \textbf{22}, 27175--27182 (2014).

\bibitem{su2017inverse}
L. Su, A.Y. Piggott, N.V. Sapra, J. Petykiewicz and J. Vuckovic, Inverse design
  and demonstration of a compact on-chip narrowband three-channel wavelength
  demultiplexer. \textit{Acs Photonics} \textbf{5}, 301--305 (2017).

\bibitem{dory2019inverse}
C. Dory, D. Vercruysse, K.Y. Yang, N.V. Sapra, A.E. Rugar, S. Sun, D.M. Lukin,
  A.Y. Piggott, J.L. Zhang, M. Radulaski and  others, Inverse-designed diamond
  photonics. \textit{Nature communications} \textbf{10}, 1--7 (2019).

\bibitem{sapra2020chip}
N.V. Sapra, K.Y. Yang, D. Vercruysse, K.J. Leedle, D.S. Black, R.J. England, L.
  Su, Y. Miao, O. Solgaard, R.L. Byer and  others, On-chip integrated
  laser-driven particle accelerator. \textit{Science} \textbf{367}, 79 (2020).

\bibitem{sheeran2000checking}
M. Sheeran, S. Singh and G. St{\aa}lmarck, Checking safety properties using
  induction and a SAT-solver. \textit{International conference on formal
  methods in computer-aided design} 127--144 (2000).

\bibitem{saeedi2013synthesis}
M. Saeedi and I.L. Markov, Synthesis and optimization of reversible circuits -
  a survey. \textit{ACM Computing Surveys (CSUR)} \textbf{45}, 21 (2013).

\bibitem{dawson2005solovay}
C.M. Dawson and M.A. Nielsen, The solovay-kitaev algorithm.
  \textit{quant-ph/0505030} (2005).

\bibitem{maslov2008quantum}
D. Maslov, G.W. Dueck, D.M. Miller and C. Negrevergne, Quantum circuit
  simplification and level compaction. \textit{IEEE Transactions on
  Computer-Aided Design of Integrated Circuits and Systems} \textbf{27},
  436--444 (2008).

\bibitem{martinez2016compiling}
E.A. Martinez, T. Monz, D. Nigg, P. Schindler and R. Blatt, Compiling quantum
  algorithms for architectures with multi-qubit gates. \textit{New Journal of
  Physics} \textbf{18}, 063029 (2016).

\bibitem{maslov2017basic}
D. Maslov, Basic circuit compilation techniques for an ion-trap quantum
  machine. \textit{New Journal of Physics} \textbf{19}, 023035 (2017).

\bibitem{mcclean2018barren}
J.R. McClean, S. Boixo, V.N. Smelyanskiy, R. Babbush and H. Neven, Barren
  plateaus in quantum neural network training landscapes. \textit{Nature
  communications} \textbf{9}, 4812 (2018).

\bibitem{brakensiek2019resolution}
J. Brakensiek, M. Heule and J. Mackey, The Resolution of Keller's Conjecture.
  \textit{arXiv:1910.03740} (2019).

\bibitem{virshup2013stochastic}
A.M. Virshup, J. Contreras-Garc{\'\i}a, P. Wipf, W. Yang and D.N. Beratan,
  Stochastic voyages into uncharted chemical space produce a representative
  library of all possible drug-like compounds. \textit{Journal of the American
  Chemical Society} \textbf{135}, 7296--7303 (2013).

\bibitem{coello2007evolutionary}
C.A.C. Coello, G.B. Lamont, D.A. Van~Veldhuizen and  others, Evolutionary
  algorithms for solving multi-objective problems. (Springer, 2007).

\bibitem{coello2010list}
C.A.C. Coello, List of references on evolutionary multiobjective optimization.
  \textit{http://www.lania.mx/ccoello/EMOO/EMOObib.html} (June 2017).

\bibitem{pan2012multiphoton}
J.W. Pan, Z.B. Chen, C.Y. Lu, H. Weinfurter, A. Zeilinger and M. {\.Z}ukowski,
  Multiphoton entanglement and interferometry. \textit{Reviews of Modern
  Physics} \textbf{84}, 777 (2012).

\bibitem{flamini2018photonic}
F. Flamini, N. Spagnolo and F. Sciarrino, Photonic quantum information
  processing: a review. \textit{Reports on Progress in Physics} \textbf{82},
  016001 (2018).

\bibitem{graffitti2017pure}
F. Graffitti, D. Kundys, D.T. Reid, A.M. Bra{\'n}czyk and A. Fedrizzi, Pure
  down-conversion photons through sub-coherence-length domain engineering.
  \textit{Quantum Science and Technology} \textbf{2}, 035001 (2017).

\bibitem{lenzini2017active}
F. Lenzini, B. Haylock, J.C. Loredo, R.A. Abrah{\~a}o, N.A. Zakaria, S.
  Kasture, I. Sagnes, A. Lemaitre, H.P. Phan, D.V. Dao and  others, Active
  demultiplexing of single photons from a solid-state source. \textit{Laser \&
  Photonics Reviews} \textbf{11}, 1600297 (2017).

\bibitem{wang201818}
X.L. Wang, Y.H. Luo, H.L. Huang, M.C. Chen, Z.E. Su, C. Liu, C. Chen, W. Li,
  Y.Q. Fang, X. Jiang and  others, 18-qubit entanglement with six photons'
  three degrees of freedom. \textit{Physical Review Letters} \textbf{120},
  260502 (2018).

\bibitem{luo2019quantum}
Y.H. Luo, H.S. Zhong, M. Erhard, X.L. Wang, L.C. Peng, M. Krenn, X. Jiang, L.
  Li, N.L. Liu, C.Y. Lu and  others, Quantum Teleportation in High Dimensions.
  \textit{Physical Review Letters} \textbf{123}, 070505 (2019).

\bibitem{bornman2019ghost}
N. Bornman, M. Agnew, F. Zhu, A. Vall{\'e}s, A. Forbes and J. Leach, Ghost
  imaging using entanglement-swapped photons. \textit{npj Quantum Information}
  \textbf{5}, 1--6 (2019).

\bibitem{wang2019boson}
H. Wang, J. Qin, X. Ding, M.C. Chen, S. Chen, X. You, Y.M. He, X. Jiang, L.
  You, Z. Wang and  others, Boson Sampling with 20 Input Photons and a 60-Mode
  Interferometer in a 10**14-Dimensional Hilbert Space. \textit{Physical review
  letters} \textbf{123}, 250503 (2019).

\bibitem{hu2019experimental}
X.M. Hu, C. Zhang, B.H. Liu, Y.F. Huang, C.F. Li and G.C. Guo, Experimental
  multi-level quantum teleportation. \textit{arXiv:1904.12249} (2019).

\bibitem{bavaresco2018measurements}
J. Bavaresco, N.H. Valencia, C. Kl{\"o}ckl, M. Pivoluska, P. Erker, N. Friis,
  M. Malik and M. Huber, Measurements in two bases are sufficient for
  certifying high-dimensional entanglement. \textit{Nature Physics}
  \textbf{14}, 1032 (2018).

\bibitem{ahn2019adaptive}
D. Ahn, Y. Teo, H. Jeong, F. Bouchard, F. Hufnagel, E. Karimi, D. Koutn{\`y},
  J. {\v{R}}eh{\'a}{\v{c}}ek, Z. Hradil, G. Leuchs and  others, Adaptive
  compressive tomography with no a priori information. \textit{Physical Review
  Letters} \textbf{122}, 100404 (2019).

\bibitem{gao2019arbitrary}
X. Gao, M. Krenn, J. Kysela and A. Zeilinger, Arbitrary d-dimensional Pauli X
  Gates of a flying Qudit. \textit{Physical Review A} \textbf{99}, 023825
  (2019).

\bibitem{krenn2017quantum}
M. Krenn, X. Gu and A. Zeilinger, Quantum experiments and graphs: Multiparty
  states as coherent superpositions of perfect matchings. \textit{Physical
  Review Letters} \textbf{119}, 240403 (2017).

\bibitem{wang2015quantum}
X.L. Wang, X.D. Cai, Z.E. Su, M.C. Chen, D. Wu, L. Li, N.L. Liu, C.Y. Lu and
  J.W. Pan, Quantum teleportation of multiple degrees of freedom of a single
  photon. \textit{Nature} \textbf{518}, 516 (2015).

\bibitem{anwer2019experimental}
H. Anwer, M. Nawareg, A. Cabello and M. Bourennane, Experimental test of
  maximal tripartite nonlocality using an entangled state and local
  measurements that are maximally incompatible. \textit{Physical Review A}
  \textbf{100}, 022104 (2019).

\bibitem{leach2002measuring}
J. Leach, M.J. Padgett, S.M. Barnett, S. Franke-Arnold and J. Courtial,
  Measuring the orbital angular momentum of a single photon. \textit{Physical
  Review Letters} \textbf{88}, 257901 (2002).

\bibitem{huber2013structure}
M. Huber and J.I. Vicente, Structure of multidimensional entanglement in
  multipartite systems. \textit{Physical Review Letters} \textbf{110}, 030501
  (2013).

\bibitem{huber2013entropy}
M. Huber, M. Perarnau-Llobet and J.I. Vicente, Entropy vector formalism and the
  structure of multidimensional entanglement in multipartite systems.
  \textit{Physical Review A} \textbf{88}, 042328 (2013).

\bibitem{PhysRevA.89.024103}
J. Ryu, C. Lee, Z. Yin, R. Rahaman, D.G. Angelakis, J. Lee and M. Zukowski,
  Multisetting Greenberger-Horne-Zeilinger theorem. \textit{Physical Review A}
  \textbf{89}, 024103 (2014).

\bibitem{lawrence2014rotational}
J. Lawrence, Rotational covariance and Greenberger-Horne-Zeilinger theorems for
  three or more particles of any dimension. \textit{Physical Review A}
  \textbf{89}, 012105 (2014).

\bibitem{lawrence2019many}
J. Lawrence, Many-qutrit Mermin inequalities with three measurement bases.
  \textit{arXiv:1910.05869} (2019).

\bibitem{zou1991induced}
X. Zou, L.J. Wang and L. Mandel, Induced coherence and indistinguishability in
  optical interference. \textit{Physical Review Letters} \textbf{67}, 318
  (1991).

\bibitem{gu2019quantum2}
X. Gu, M. Erhard, A. Zeilinger and M. Krenn, Quantum experiments and graphs II:
  Quantum interference, computation, and state generation. \textit{Proceedings
  of the National Academy of Sciences} \textbf{116}, 4147--4155 (2019).

\bibitem{gu2019quantum3}
X. Gu, L. Chen, A. Zeilinger and M. Krenn, Quantum experiments and graphs. III.
  High-dimensional and multiparticle entanglement. \textit{Physical Review A}
  \textbf{99}, 032338 (2019).

\bibitem{krenn2019questions}
M. Krenn, X. Gu and D. Solt{\'e}sz, Questions on the Structure of Perfect
  Matchings inspired by Quantum Physics. \textit{Proceedings of the 2nd
  Croatian Combinatorial Days} 57--70 (2019).

\bibitem{lehman2018surprising}
J. Lehman, J. Clune, D. Misevic, C. Adami, L. Altenberg, J. Beaulieu, P.J.
  Bentley, S. Bernard, G. Beslon, D.M. Bryson and  others, The surprising
  creativity of digital evolution: A collection of anecdotes from the
  evolutionary computation and artificial life research communities.
  \textit{arXiv:1803.03453} (2018).

\bibitem{wang2019integrated}
J. Wang, F. Sciarrino, A. Laing and M.G. Thompson, Integrated photonic quantum
  technologies. \textit{Nature Photonics} 1--12 (2019).

\bibitem{fengprogress}
L.T. Feng, G.C. Guo and X.F. Ren, Progress on Integrated Quantum Photonic
  Sources with Silicon. \textit{Advanced Quantum Technologies} 1900058 .

\bibitem{reck1994experimental}
M. Reck, A. Zeilinger, H.J. Bernstein and P. Bertani, Experimental realization
  of any discrete unitary operator. \textit{Physical Review Letters}
  \textbf{73}, 58 (1994).

\bibitem{clements2016optimal}
W.R. Clements, P.C. Humphreys, B.J. Metcalf, W.S. Kolthammer and I.A. Walmsley,
  Optimal design for universal multiport interferometers. \textit{Optica}
  \textbf{3}, 1460--1465 (2016).

\bibitem{tischler2018quantum}
N. Tischler, C. Rockstuhl and K. S{\l}owik, Quantum optical realization of
  arbitrary linear transformations allowing for loss and gain. \textit{Physical
  Review X} \textbf{8}, 021017 (2018).

\bibitem{giovannetti2011advances}
V. Giovannetti, S. Lloyd and L. Maccone, Advances in quantum metrology.
  \textit{Nature photonics} \textbf{5}, 222 (2011).

\bibitem{salimans2017evolution}
T. Salimans, J. Ho, X. Chen, S. Sidor and I. Sutskever, Evolution strategies as
  a scalable alternative to reinforcement learning. \textit{arXiv:1703.03864}
  (2017).

\bibitem{sutton1998introduction}
R.S. Sutton, A.G. Barto and  others, Introduction to reinforcement learning.
  (MIT press Cambridge, 1998).

\bibitem{mnih2015human}
V. Mnih, K. Kavukcuoglu, D. Silver, A.A. Rusu, J. Veness, M.G. Bellemare, A.
  Graves, M. Riedmiller, A.K. Fidjeland, G. Ostrovski and  others, Human-level
  control through deep reinforcement learning. \textit{Nature} \textbf{518},
  529 (2015).

\bibitem{vinyals2019grandmaster}
O. Vinyals, I. Babuschkin, W.M. Czarnecki, M. Mathieu, A. Dudzik, J. Chung,
  D.H. Choi, R. Powell, T. Ewalds, P. Georgiev and  others, Grandmaster level
  in StarCraft II using multi-agent reinforcement learning. \textit{Nature}
  \textbf{575}, 350--354 (2019).

\bibitem{jaderberg2019human}
M. Jaderberg, W.M. Czarnecki, I. Dunning, L. Marris, G. Lever, A.G. Castaneda,
  C. Beattie, N.C. Rabinowitz, A.S. Morcos, A. Ruderman and  others,
  Human-level performance in 3D multiplayer games with population-based
  reinforcement learning. \textit{Science} \textbf{364}, 859--865 (2019).

\bibitem{silver2018general}
D. Silver, T. Hubert, J. Schrittwieser, I. Antonoglou, M. Lai, A. Guez, M.
  Lanctot, L. Sifre, D. Kumaran, T. Graepel and  others, A general
  reinforcement learning algorithm that masters chess, shogi, and Go through
  self-play. \textit{Science} \textbf{362}, 1140--1144 (2018).

\bibitem{schmidhuber1991curious}
J. Schmidhuber, Curious model-building control systems. \textit{Proc.
  international joint conference on neural networks} 1458--1463 (1991).

\bibitem{schmidhuber1991possibility}
J. Schmidhuber, A possibility for implementing curiosity and boredom in
  model-building neural controllers. \textit{Proc. of the international
  conference on simulation of adaptive behavior: From animals to animats}
  222--227 (1991).

\bibitem{schmidhuber2010formal}
J. Schmidhuber, Formal theory of creativity, fun, and intrinsic motivation
  (1990--2010). \textit{IEEE Transactions on Autonomous Mental Development}
  \textbf{2}, 230--247 (2010).

\bibitem{pathak2017curiosity}
D. Pathak, P. Agrawal, A.A. Efros and T. Darrell, Curiosity-driven exploration
  by self-supervised prediction. \textit{Proceedings of the IEEE Conference on
  Computer Vision and Pattern Recognition Workshops} 16--17 (2017).

\bibitem{pathak18largescale}
Y. Burda, H. Edwards, D. Pathak, A. Storkey, T. Darrell and A.A. Efros,
  Large-Scale Study of Curiosity-Driven Learning. \textit{ICLR} (2019).

\bibitem{briegel2012projective}
H.J. Briegel and G. Cuevas, Projective simulation for artificial intelligence.
  \textit{Scientific reports} \textbf{2}, 400 (2012).

\bibitem{briegel2012creative}
H.J. Briegel, On creative machines and the physical origins of freedom.
  \textit{Scientific reports} \textbf{2}, 522 (2012).

\bibitem{hochreiter1997long}
S. Hochreiter and J. Schmidhuber, Long short-term memory. \textit{Neural
  computation} \textbf{9}, 1735--1780 (1997).

\bibitem{qiang2018large}
X. Qiang, X. Zhou, J. Wang, C.M. Wilkes, T. Loke, S. O'Gara, L. Kling, G.D.
  Marshall, R. Santagati, T.C. Ralph and  others, Large-scale silicon quantum
  photonics implementing arbitrary two-qubit processing. \textit{Nature
  photonics} \textbf{12}, 534 (2018).

\bibitem{wang2018multidimensional}
J. Wang, S. Paesani, Y. Ding, R. Santagati, P. Skrzypczyk, A. Salavrakos, J.
  Tura, R. Augusiak, L. Man{\v{c}}inska, D. Bacco and  others, Multidimensional
  quantum entanglement with large-scale integrated optics. \textit{Science}
  \textbf{360}, 285--291 (2018).

\bibitem{lu2019three}
L. Lu, L. Xia, Z. Chen, L. Chen, T. Yu, T. Tao, W. Ma, Y. Pan, X. Cai, Y. Lu
  and  others, Three-dimensional entanglement on a silicon chip.
  \textit{arXiv:1911.08807} (2019).

\bibitem{weedbrook2012gaussian}
C. Weedbrook, S. Pirandola, R. Garc{\'\i}a-Patr{\'o}n, N.J. Cerf, T.C. Ralph,
  J.H. Shapiro and S. Lloyd, Gaussian quantum information. \textit{Reviews of
  Modern Physics} \textbf{84}, 621 (2012).

\bibitem{lenzini2018integrated}
F. Lenzini, J. Janousek, O. Thearle, M. Villa, B. Haylock, S. Kasture, L. Cui,
  H.P. Phan, D.V. Dao, H. Yonezawa and  others, Integrated photonic platform
  for quantum information with continuous variables. \textit{Science advances}
  \textbf{4}, eaat9331 (2018).

\bibitem{zhang2019integrated}
G. Zhang, J. Haw, H. Cai, F. Xu, S. Assad, J. Fitzsimons, X. Zhou, Y. Zhang, S.
  Yu, J. Wu and  others, An integrated silicon photonic chip platform for
  continuous-variable quantum key distribution. \textit{Nature Photonics}
  \textbf{13}, 839--842 (2019).

\bibitem{killoran2018continuous}
N. Killoran, T.R. Bromley, J.M. Arrazola, M. Schuld, N. Quesada and S. Lloyd,
  Continuous-variable quantum neural networks. \textit{Physical Review
  Research} \textbf{1}, 033063 (2019).

\bibitem{menicucci2014fault}
N.C. Menicucci, Fault-tolerant measurement-based quantum computing with
  continuous-variable cluster states. \textit{Physical Review Letters}
  \textbf{112}, 120504 (2014).

\bibitem{killoran2019strawberry}
N. Killoran, J. Izaac, N. Quesada, V. Bergholm, M. Amy and C. Weedbrook,
  Strawberry fields: A software platform for photonic quantum computing.
  \textit{Quantum} \textbf{3}, 129 (2019).

\bibitem{sabapathy2019production}
K.K. Sabapathy, H. Qi, J. Izaac and C. Weedbrook, Production of photonic
  universal quantum gates enhanced by machine learning. \textit{Physical Review
  A} \textbf{100}, 012326 (2019).

\bibitem{morizur2010programmable}
J.F. Morizur, L. Nicholls, P. Jian, S. Armstrong, N. Treps, B. Hage, M. Hsu, W.
  Bowen, J. Janousek and H.A. Bachor, Programmable unitary spatial mode
  manipulation. \textit{JOSA A} \textbf{27}, 2524--2531 (2010).

\bibitem{fontaine2019laguerre}
N.K. Fontaine, R. Ryf, H. Chen, D.T. Neilson, K. Kim and J. Carpenter,
  Laguerre-Gaussian mode sorter. \textit{Nature communications} \textbf{10},
  1865 (2019).

\bibitem{brandt2019high}
F. Brandt, M. Hiekkam{\"a}ki, F. Bouchard, M. Huber and R. Fickler,
  High-dimensional quantum gates using full-field spatial modes of photons.
  \textit{arXiv:1907.13002} (2019).

\bibitem{rotter2017light}
S. Rotter and S. Gigan, Light fields in complex media: Mesoscopic scattering
  meets wave control. \textit{Reviews of Modern Physics} \textbf{89}, 015005
  (2017).

\bibitem{fickler2017custom}
R. Fickler, M. Ginoya and R.W. Boyd, Custom-tailored spatial mode sorting by
  controlled random scattering. \textit{Physical Review B} \textbf{95}, 161108
  (2017).

\bibitem{leedumrongwatthanakun2019programmable}
S. Leedumrongwatthanakun, L. Innocenti, H. Defienne, T. Juffmann, A. Ferraro,
  M. Paternostro and S. Gigan, Programmable linear quantum networks with a
  multimode fibre. \textit{Nature Photonics} 1--4 (2019).

\bibitem{krenn2019selfies}
M. Krenn, F. H{\"a}se, A. Nigam, P. Friederich and A. Aspuru-Guzik, SELFIES: a
  robust representation of semantically constrained graphs with an example
  application in chemistry. \textit{arXiv:1905.13741} (2019).

\bibitem{heule2016solving}
M.J. Heule, O. Kullmann and V.W. Marek, Solving and verifying the boolean
  pythagorean triples problem via cube-and-conquer. \textit{International
  Conference on Theory and Applications of Satisfiability Testing} 228--245
  (2016).

\bibitem{heule2017science}
M.J. Heule and O. Kullmann, The science of brute force.. \textit{Commun. ACM}
  \textbf{60}, 70--79 (2017).

\bibitem{higgins2017beta}
I. Higgins, L. Matthey, A. Pal, C. Burgess, X. Glorot, M. Botvinick, S. Mohamed
  and A. Lerchner, beta-VAE: Learning Basic Visual Concepts with a Constrained
  Variational Framework.. \textit{ICLR} \textbf{2}, 6 (2017).

\bibitem{chen2018isolating}
T.Q. Chen, X. Li, R.B. Grosse and D.K. Duvenaud, Isolating sources of
  disentanglement in variational autoencoders. \textit{Advances in Neural
  Information Processing Systems} 2610--2620 (2018).

\bibitem{lusch2018deep}
B. Lusch, J.N. Kutz and S.L. Brunton, Deep learning for universal linear
  embeddings of nonlinear dynamics. \textit{Nature communications} \textbf{9},
  4950 (2018).

\bibitem{iten2020discovering}
R. Iten, T. Metger, H. Wilming, L. Del~Rio and R. Renner, Discovering physical
  concepts with neural networks. \textit{Physical Review Letters} \textbf{124},
  010508 (2020).

\bibitem{Nautrup2020operationally}
H.P. Nautrup, T. Metger, R. Iten, S. Jerbi, L.M. Trenkwalder, H. Wilming, H.J.
  Briegel and R. Renato, Operationally meaningful representations of physical
  systems in neural networks. \textit{arXiv:2001.00593} (2020).

\bibitem{bentley2002introduction}
P.J. Bentley and D.W. Corne, An introduction to creative evolutionary systems.
  \textit{Creative evolutionary systems} 1--75 (2002).

\bibitem{pavivcic2019automated}
M. Pavi{\v{c}}i{\'c}, M. Waegell, N.D. Megill and P. Aravind, Automated
  generation of Kochen-Specker sets. \textit{Scientific reports} \textbf{9},
  6765 (2019).

\bibitem{goyeneche2015absolutely}
D. Goyeneche, D. Alsina, J.I. Latorre, A. Riera and K. {\.Z}yczkowski,
  Absolutely maximally entangled states, combinatorial designs, and
  multiunitary matrices. \textit{Physical Review A} \textbf{92}, 032316 (2015).

\bibitem{bengtsson2017geometry}
I. Bengtsson and K. {\.Z}yczkowski, Geometry of quantum states: an introduction
  to quantum entanglement. (Cambridge university press, 2017).

\bibitem{horodecki2020five}
P. Horodecki, L. Rudnicki and K. Zyczkowski, Five open problems in quantum
  information. \textit{arXiv:2002.03233} (2020).

\end{thebibliography}
\let\addcontentsline\oldaddcontentsline

\end{document}